\documentclass{article}
\usepackage{jheppub} 
\usepackage{graphicx} 
\usepackage{amsmath,amsfonts,amssymb,array,braket,booktabs,empheq,lmodern,float,mathtools,nccmath,physics,slashed,bm,tikz,simpler-wick}
\usetikzlibrary{shapes.misc,shapes.geometric}
\usepackage{color}
\usepackage{wrapfig}
\usepackage{comment}
\usepackage{bbm}

\newcommand{\beq}{\begin{equation}}
\newcommand{\eeq}{\end{equation}}
\newcommand{\beqn}{\begin{eqnarray}}
\newcommand{\eeqn}{\end{eqnarray}}

\newcommand{\cN}{\mathcal{N}}

\newcommand{\bdelta}{\bar{\delta}}

\title{Wigner negativity, random matrices and gravity}

\author[a,1]{Ritam Basu,\note{ritam.basu@tifr.res.in}}
\author[b,2]{Pratyusha Chowdhury,\note{P.Chowdhury@soton.ac.uk}}
\author[c,3]{ Anirban Ganguly,\note{anirbang@cmi.ac.in}}
\author[a,4]{ Souparna Nath,\note{souparna.nath@tifr.res.in}}
\author[a,5]{ Onkar Parrikar\note{parrikar@theory.tifr.res.in}}
\author[a,6]{Suprakash Paul \note{suprakash.paul@tifr.res.in}}

\affiliation[a]{Department of Theoretical Physics, 
Tata Institute of Fundamental Research, 1 Homi Bhabha Road, Mumbai 400005, India}
\affiliation[b]{School of Physics \& Astronomy and STAG Research Centre, University of Southampton, Highfield, Southampton SO17 IBJ, United Kingdom}
\affiliation[c]{Chennai Mathematical Institute, Plot H1, SIPCOT IT Park, Padur PO, Siruseri, 603103 India}

\abstract{
Given a choice of an ordered, orthonormal basis for a $D$-dimensional Hilbert space, one can define a discrete version of the Wigner function — a quasi-probability distribution which represents any quantum state as a real, normalized function on a discrete phase space. The Wigner function, in general, takes on negative values, and the amount of negativity in the Wigner function gives an operationally meaningful measure of the complexity of simulating the quantum state on a classical computer. Further, Wigner negativity also gives a lower bound on an entropic measure of spread complexity. In this paper, we study the growth of Wigner negativity for a generic initial state under time evolution with chaotic Hamiltonians. In \cite{2024JHEP...05..264B}, a perturbative argument was given to show that the Krylov basis minimizes the early time growth of Wigner negativity in the large-$D$ limit. Using tools from random matrix theory, here we show that for a generic choice of basis, the Wigner negativity for a classical initial state becomes exponentially large in an $O(1)$ amount of time evolution. On the other hand, we show that in the Krylov basis the negativity grows at most as a power law, and becomes exponentially large only at exponential times. We take this as evidence that the Krylov basis is ideally suited for a dual, semi-classical effective description of chaotic quantum dynamics for large-$D$ at sub-exponential times. For the Gaussian unitary ensemble, this effective description is the $q\to 0$ limit of $q$-deformed JT gravity. 
}

\begin{document}
\maketitle
\flushbottom

\parskip=10pt
\newpage

\section{Introduction}
The AdS/CFT correspondence \cite{Maldacena:1997re, Witten:1998qj} gives a map between a theory of quantum gravity in asymptotically AdS spacetime and a more conventional quantum theory which lives on the one-lower dimensional boundary of AdS. Remarkably, the duality allows us to compute certain quantities in the boundary quantum theory at large  $N$ (where $N$ is, loosely speaking, some measure of the number of degrees of freedom) and strong coupling via classical, gravitational calculations in the bulk. One may say that gravity provides the most efficient set of variables to describe aspects of the strong-coupling dynamics in the dual quantum mechanical theory at large $N$. On the other hand, the boundary theory gives an ultra-violet completion for gravity. Given the success of the AdS/CFT correspondence, it is natural to look for a deeper explanation for how AdS/CFT works. Here, we look for such an explanation from a Hilbert space point of view, i.e., given a large-$N$, strongly coupled quantum mechanical system, how do we know whether it contains some version of gravity within its Hilbert space? (see also \cite{Gopakumar:2003ns, Gopakumar:2004qb, Gopakumar:2005fx, Gopakumar:2011ev, Gopakumar:2022djw} for a different approach involving open-closed string duality).

In \cite{2024JHEP...05..264B}, it was suggested that the stabilizer formalism from the theory of quantum computation \cite{Veitch_2012, Veitch_2014} could provide a useful framework to formulate the above question. One of the landmark results in the stabilizer formalism is the Gottesman-Knill theorem \cite{gottesman1998heisenberg, Aaronson:2004xuh, Mari_2012, Pashayan_2015, Wang_2019}, which identifies a large class of quantum circuits that can efficiently be simulated on a classical computer. An essential role in this analysis is played by the discrete Wigner function \cite{Wigner:1932, WOOTTERS19871, Leonard, sphere, quantumcomp, Galois, Gross_2006, classicality}, a quasi-probability distribution that can be assigned to any quantum state, and more generally to any quantum operation. While the discrete Wigner function mimics a probability distribution on a discrete ``phase space'', one essential point of departure is that it is not always everywhere non-negative for all states -- had this been the case, quantum mechanics would reduce to classical probability theory. The Gottesman-Knill theorem, loosely speaking, states that any quantum circuit involving quantum operations with non-negative Wigner functions (i.e., whose Wigner functions are genuine probability distributions) can be efficiently simulated on a classical computer. Furthermore, the amount of negativity in the discrete Wigner function representation of intermediate states provides an operationally meaningful measure of the complexity of classically simulating a quantum circuit \cite{Veitch_2012, Veitch_2014, Pashayan_2015}. This quantity, which we will refer to as the \emph{Wigner negativity} (see equation \eqref{negdef} for the definition), will be the central focus of this paper. For the reasons explained above, Wigner negativity is thought of in the quantum information literature as a measure of an inherently quantum resource called ``magic'' that makes quantum computation essentially different and more powerful as compared to classical computation.    

A crucial point is that the discrete Wigner function and its negativity is defined with respect to a choice of computational basis. We can think of the Wigner negativity as an abstract way of characterizing the efficiency of describing quantum dynamics in a given choice of basis. In \cite{2024JHEP...05..264B}, the authors studied the following question: given an initial state $\psi_0$ and the time evolution operator $U(t) = e^{-itH}$, can we find a computational basis such that the growth of Wigner negativity is minimized with respect to this basis? If such a basis exists, then it would correspond to a convenient choice of variables in terms of which the quantum time evolution of $\psi_0$ looks as classical as possible. In \cite{2024JHEP...05..264B}, a perturbative argument was given to show that a basis which does the job is the \emph{Krylov basis} \cite{Balasubramanian:2022tpr, Balasubramanian:2022dnj}. The Krylov basis is an ordered, orthonormal basis obtained by applying the Gram-Schmidt procedure to the set of states $\{\psi_0, H\psi_0, H^2\psi_0,\cdots\}$. This basis has appeared in recent discussions of operator growth, chaos and spread complexity \cite{Parker:2018yvk, Rabinovici_2021, caputa2021geometry, Balasubramanian:2024ghv} (see \cite{Nandy:2024htc, Baiguera:2025dkc} for more details and references). As we will explain in section \ref{sec:prelim}, these various notions of complexity are not unrelated; in particular, Wigner negativity gives a lower bound on a certain entropic measure of spread complexity.

The goal of the present paper is to study finite time evolution of Wigner negativity for chaotic Hamiltonians in the limit where the Hilbert space dimension becomes large (see \cite{Goto, Leone2021quantumchaosis} for previous work on Wigner negativity and chaos). We can think of this setup as a 0+1-dimensional toy model for the strongly coupled, large $N$ dynamics of holographic conformal field theories. Our main results are as follows:

1. For a generic choice of computational basis, the Wigner negativity for a classical initial state grows very rapidly and saturates to an exponentially large value within an $O(1)$ amount of time evolution, where by exponentially large we mean scaling exponentially with the log of the Hilbert space dimension. Thus, the Wigner negativity does not admit a good large $D$ limit, even at small times.

2. In the Krylov basis, the Wigner negativity at large $O(1)$ times grows as a power law and thus remains finite for any $O(1)$ amount of time evolution in the $D\to \infty$ limit. Thus, the Krylov basis gives a good effective, semi-classical description of chaotic quantum dynamics at sub-exponential times. However, this effective description breaks down at an exponentially large time scale, where the negativity saturates to an exponentially large value.

Of course, the negativity is difficult to compute analytically for a specific choice of a chaotic Hamiltonian, so in order to make progress analytically, we resort to tools from random matrix theory. We take the Hamiltonian to be a random matrix drawn from the Gaussian unitary ensemble (GUE), and calculate the ensemble averaged Wigner negativity. One key point (especially relevant for the derivation of the first result above) is that even at small times, the ensemble average of the negativity of the Wigner function is not the same as the negativity of the averaged Wigner function. Furthermore, the negativity involves a non-analytic function (namely, the absolute value of the Wigner function). We use the replica trick to deal with these complications. The output of this calculation is an explicit formula for the ensemble-average of the Wigner negativity (with respect to a generic choice of basis) in terms of the survival amplitude or equivalently, the spectral form factor. 

The calculation of Wigner negativity in the Krylov basis is facilitated by the fact that in the large-$D$ limit and for $O(1)$ times, the dynamics is controlled by an effective Hamiltonian, which turns out to be solvable. The fact that the negativity only grows gradually in the Krylov basis -- and, in particular, admits a good large $D$ limit at $O(1)$ times -- suggests that this basis provides an efficient choice of semi-classical variables to describe chaotic quantum dynamics. Interestingly, the effective theory in the Krylov basis precisely coincides with the $q\to 0 $ limit of $q$-deformed JT gravity \cite{Berkooz:2018jqr, Berkooz:2022mfk, Lin:2022rbf, Jafferis:2022wez, Rabinovici:2023yex, Okuyama:2022szh, Okuyama:2023byh, Nandy:2024zcd, Almheiri:2024xtw,Xu:2024gfm, Xu:2024hoc, Blommaert:2024ydx, Blommaert:2025avl, Miyaji:2025ucp}, i.e., the theory of chords which is dual to the double-scaled SYK model in the large-$N$ limit. However, numerical calculations show that the negativity saturates to an exponentially large value at an exponential time-scale. This gives a sharp diagnostic of the breakdown of the effective description. We currently do not have an analytical understanding of these late time features. 

The rest of the paper is organized as follows: in section \ref{sec:prelim}, we review the necessary background and discuss how Wigner negativity is related to classical simulation complexity, and to entropic measures of spread complexity. In section \ref{sec:generic}, we compute the ensemble average of the Wigner negativity with respect to a generic choice of basis. In section \ref{sec:Krylov}, we describe the effective Hamiltonian in the Krylov basis and use this effective description to argue that the Wigner negativity cannot grow faster than $O(\sqrt{t})$. We end with some open questions and future directions in section \ref{sec:discussion}.

\section{Preliminaries}\label{sec:prelim}
\subsection{The discrete Wigner function}

For any finite dimensional quantum system of odd prime dimension $D$,\footnote{The discrete Wigner function formalism is cleanest when $D$ is an odd prime number, although almost all of the results work more generally for odd $D$, and generalizations exist for even $D$ as well. For simplicity, we will take $D$ to be an odd prime number in this work.} there exists a natural quasi-probability representation called the \emph{discrete Wigner function} \cite{WOOTTERS19871, Leonard, sphere, quantumcomp, Galois, Gross_2006, classicality}. Let $\left\{|k\rangle\right\}_{k=0}^{D-1}$ be an ordered, orthonormal basis for the Hilbert space. The essential idea is to interpret this as the ``position'' basis, and correspondingly construct a ``phase space''. We define the \emph{discrete phase space} $\mathcal{P}$ as the lattice $\mathbb{Z}_D \times \mathbb{Z}_D$ of size $D^2$. 
With respect to the chosen basis, we define a set of $D^2$ operators $A(q,p)$ called phase-point operators, each labelled by a phase space point  \cite{WOOTTERS19871}:
\begin{equation}\label{eq:A_Wooters}
A(q,p) = \sum_{k,\ell=0}^{D-1} \widehat{\delta}_{2q,k+l}e^{\frac{2\pi i}{D} (k-\ell)p } \ketbra{k}{\ell}.
\end{equation}
Note that the hatted Kronecker delta $\widehat{\delta}$ is the $\text{mod}\,D$ version, i.e., it is one when $(k+\ell) = 2q\,\text{mod}\,D$, and zero otherwise. We will often use the notation $\vec{\alpha} = (q,p)$ to denote a phase-space point, and correspondingly denote the phase-point operator as $A_{\vec{\alpha}}$. These phase-point operators satisfy the following properties:
\begin{equation}\label{eq:A_properties}
\begin{split}
&\Tr(A_{\vec{\alpha}}) = 1,\\
&\Tr(A_{\vec{\alpha}}A_{\vec{\beta}}) = D\,\delta_{\vec{\alpha},\vec{\beta}},\\
&\frac{1}{D} \sum_{\vec{\alpha}}A_{\vec{\alpha}} = \mathbf{1}.
\end{split}
\end{equation}
The \emph{discrete Wigner function} for a density matrix $\rho$ is now defined as:
\beq 
W_{\rho}(q,p) = \frac{1}{D} \mathrm{Tr}\left(\rho A(q,p)\right).
\eeq 
The Wigner function should be thought of as an attempt to represent the quantum state $\rho$ as a probability distribution in phase space, as would be the case in classical mechanics. Indeed, it satisfies the following properties:
\begin{enumerate}
\item The discrete Wigner function is real and is unit normalized, i.e.,
\begin{equation}\label{eq:Wigner_norm}
\Tr(\rho) = 1 \implies \sum_{{\vec{\alpha}}} W_{\rho}(\vec{\alpha}) = 1.
\end{equation}
\item Summing over one of the directions, say either $p$ or $q$, reduces the Wigner function to the probability density along the other direction:
\beq 
\sum_{p=0}^{D-1} W_{\rho}(q,p) = \langle q|\rho|q\rangle,\;\;\;\sum_{q=0}^{D-1} W_{\rho}(q,p) = \langle p|\rho|p\rangle,
\eeq 
where we have defined the ``momentum eigenstates'' as $|p\rangle = \frac{1}{\sqrt{D}}\sum_{q=0}^{D-1}e^{\frac{2\pi i pq}{D}}|q\rangle$. It is an easy exercise to check that we could have equivalently defined the Wigner function in terms of the momentum basis. 
\item The time evolution of the discrete Wigner function follows a discrete version of the Moyal equation:
\beq \label{dis_Moyal}
\frac{dW_{\rho}(\vec{\alpha})}{dt} = \frac{2}{ D}\sum_{\vec{\beta},\vec{\gamma}}\sin\left(\frac{4\pi}{D}\mathcal{A}_{\vec{\alpha}\vec{\beta}\vec{\gamma}}\right)H(\vec{\beta})W_{\rho}(\vec{\gamma}),
\eeq 
where 
\beq 
\mathcal{A}_{\vec{\alpha}\vec{\beta}\vec{\gamma}} = \alpha_2(\gamma_1-\beta_1) + \beta_2(\alpha_1-\gamma_1) + \gamma_2 (\beta_1-\alpha_1),
\eeq 
and the Hamiltonian is expressed in the basis of the phase point operators as $H = \displaystyle\sum_{\vec{\alpha}} H(\vec{\alpha}) A_{\vec{\alpha}}$. The discrete Moyal equation can be thought of as the quantum analog of the Liouville equation from classical mechanics.  
\end{enumerate}
While the above properties seem to suggest that we should regard the Wigner function as a probability distribution in phase space, this interpretation fails for an important reason -- the Wigner function can take negative values at some points in phase space. Indeed, this is why quantum mechanics is essentially different from classical probabilistic dynamics. The negativity of the Wigner function will play a central role in this work. 

So far we have discussed Wigner functions for states, but Wigner functions can also be defined for quantum operations/channels \cite{Mari_2012, Wang_2019}. For a quantum channel $\mathcal{E}(\rho) = \sum_m E_m\,\rho\,E_m^{\dagger}$ given in terms of a Krauss operator representation $\{E_m\}$, the Wigner function is defined as 
\beq 
W_{\mathcal{E}}(\vec{\beta}|\vec{\alpha})  = \frac{1}{D}\sum_m\mathrm{Tr}\left(A_{\vec{\beta}}\,E_m\,A_{\vec{\alpha}}\,E_m^{\dagger}\right).
\eeq 
This can be thought of as a representation of the quantum channel as a quasi-stochastic matrix on phase space. Indeed, given an input state $\rho$ with the Wigner function $W_{\rho}(\vec{\alpha})$, the Wigner function of the output state $\mathcal{E}(\rho)$ satisfies
\beq 
W_{\mathcal{E}(\rho)}(\vec{\beta}) = \sum_{\vec{\alpha}} W_{\mathcal{E}}(\vec{\beta}|\vec{\alpha}) \,W_{\rho}(\vec{\alpha}).
\eeq 
Similarly, for a positive operator valued measure $\{M_n\}$, the corresponding Wigner function is defined as:
\beq 
W(n | \vec{\alpha}) = \mathrm{Tr}\left(M_n\,A_{\vec{\alpha}}\right).
\eeq 
This definition is designed to satisfy:
\beq 
\text{Tr}\,(M_n\rho) = \sum_{\vec{\alpha}} W(n| \vec{\alpha})W_{\rho}(\vec{\alpha}),
\eeq 
and turns the POVM into a quasi conditional probability distribution. In this way, the outcome of a general quantum circuit:
\beq
P_n = \mathrm{Tr}\left(M_n\,\mathcal{E}_m \cdots \mathcal{E}_1(\rho)\right),
\eeq 
translates to quasi-stochastic evolution in terms of the Wigner function representations:
\beq \label{outcome}
P_n = \sum_{\vec{\alpha}}\sum_{\vec{\beta_1}\cdots,\vec{\beta}_m}\sum_{\vec{\gamma}}W_{M_n}(\vec{\alpha})\, W_{\mathcal{E}_m}(\vec{\beta_m}|\vec{\beta}_{m-1})\cdots W_{\mathcal{E}_1}(\vec{\beta_1}|\vec{\gamma})\,W_{\rho}(\vec{\gamma}),
\eeq
with the important caveat that the Wigner functions are not always everywhere positive. From the above discussion, it is natural to regard the set of states, operations and measurements with non-negative Wigner functions as being \emph{classical}, and Wigner negativity as a measure of non-classicality. One can make several arguments for this; we will present two arguments below, and mention one more in passing.

\subsection{Wigner negativity as complexity of classical simulation}

A direct link between positivity of the Wigner function and classicality is provided by the \emph{Gottesman-Knill theorem} \cite{gottesman1998heisenberg, Aaronson:2004xuh} -- any quantum circuit which involves initializing to a Wigner positive state, quantum operations with positive Wigner functions, and measurements defined by POVMs with positive Wigner functions can be efficiently simulated on a classical computer \cite{Mari_2012, Veitch_2012}. The essential idea is that any quantum circuit constructed out of Wigner positive elements 
 reduces to classical stochastic  (Markovian) evolution in terms of the Wigner function representation; this follows from equation \eqref{outcome}, where now all the Wigner functions are genuine probability distributions. As long as each of these Wigner probability distributions can be efficiently sampled,\footnote{Everywhere positive Wigner functions are severely constrained and have very few free parameters, so we expect that sampling from such distributions should be efficient.} then we can also efficiently sample the output probability distribution of the quantum circuit on a classical computer. Interestingly, one can generalize this argument to the case where circuit elements are not Wigner positive, and one finds that the more the Wigner negativity in the circuit, the less efficient the classical simulation \cite{Pashayan_2015, Wang_2019}.

From the above discussion, it is natural to interpret the negativity of the Wigner function as a measure of the complexity of classical simulation. One way to make this more concrete is in terms of the resource theory of stabilizer computation. Pure states with positive Wigner functions -- also called \emph{stabilizer states} -- are very highly constrained; for instance, any pure state with a positive Wigner function necessarily has a Gaussian wavefunction \cite{Gross_2006}. A mixed state is called stabilizer if it can be written as a convex combination of pure stabilizer states. \emph{Clifford unitaries} constitute a special subgroup of the unitary group which map stabilizer states to stabilizer states. These are also very highly constrained, and can be characterized completely in terms of sympelctic affine transformations on the discrete phase space. So, it is clear that circuits which only involve stabilizer states and Clifford unitaries only evolve within a very small subset of states in the Hilbert space. More generally, one is also allowed to bring in ancillas initialized to stabilizer states, perform joint Clifford unitaries, make measurements on the ancilla in the computational basis, use classical conditioning etc.; all these operations constitute ``allowed'' operations in \emph{stabilizer protocols}. However, stabilizer circuits are \emph{not} universal -- one can think of them as a class of circuits that can be efficiently simulated classically. In order to achieve universal quantum computation, one needs an additional resource, sometimes referred to as ``magic'' \cite{Bravyi_2005}. A state is said to be \emph{magic} if it is not a stabilizer state. Much like entanglement plays the role of a resource in the theory of quantum communication with local quantum operations and classical communication being treated as ``cheap'', \emph{magic} plays the role of a resource in the theory of quantum computation, with stabilizer states and operations being treated as cheap \cite{Veitch_2014}.  

In order to quantify the notion of magic, one defines a \emph{magic monotone} $\mathcal{M}$ as any real-valued, non-negative function on density matrices which is, on average, non-increasing under any stabilizer protocol \cite{Veitch_2014}. By on average, we mean that if $\Lambda$ is a stabilizer protocol potentially involving a measurement, and if under $\Lambda$ we have $\rho \stackrel{\Lambda}{\to} \{p_i,\sigma_i\}$ (i.e., the outcome is the state $\sigma_i$ with probability $p_i$), then 
\beq 
\mathcal{M}(\rho) \geq \sum_i p_i \mathcal{M}(\sigma_i).
\eeq 
Importantly, it may be the case that for some $i$, $\mathcal{M}(\rho) < \mathcal{M}(\rho_i)$, as long as the corresponding probability is sufficiently small. It turns out that the \emph{negativity} of the Wigner function, defined as:\footnote{It is more accurate to define $s\mathcal{N}(\psi) = \frac{1}{2}(\sum_{q,p} |W_{\psi}(q,p)| - 1)$ as the negativity. This quantity was referred to as the sum negativity in \cite{Veitch_2014}, and vanishes on stabilizer states. However, we will use the convention in equation \eqref{negdef} in this paper.}
\begin{equation}
    \label{negdef}
    \cN(\rho)= \sum_{q,p} |W_{\rho}(q,p)|,
\end{equation}
is a magic monotone \cite{Veitch_2014}. This gives the Wigner negativity a direct operational meaning: assuming there exists a stabilizer protocol which converts an initial state $\rho_0$ to a final state $\rho$ with probability $p$, then 
\beq 
p \leq \frac{\mathcal{N}(\rho_0)}{\mathcal{N}(\rho)}.
\eeq 
Thus, in order to obtain the desired state, one would need to repeat \emph{any} such protocol at least $\frac{\mathcal{N}(\rho)}{\mathcal{N}(\rho_0)}$ number of times. In this sense, one may regard Wigner negativity as a measure of \emph{stabilizer complexity}. Another magic monotone that is closely related to the negativity is the \emph{mana}, defined as $\mathcal{M}(\rho) := \log\,\mathcal{N}(\rho).$ The mana has the advantage that it is additive under tensor products.  

In the rest of this work, we will study the negativity of the Wigner function $\mathcal{N}(\rho)$ as a measure of the non-classicality of a state. As reviewed above, the resource theory of stabilizer quantum computation provides a concrete interpretation of the Wigner negativity as a measure of the complexity of classical simulation. The negativity has several nice properties. It is evident from the definition that $\mathcal{N} \geq 1$, where the lower bound is saturated if and only if the Wigner function is nowhere negative. The negativity is also bounded above by $
\sqrt{D}$. To see this, we note from equation \eqref{eq:A_properties} that
\begin{equation}
\sum_{\vec{\alpha}} W_{\vec{\alpha}}^2 = \frac{1}{D}\Tr(\rho^2). 
\end{equation}
Together with Jensen's inequality and the fact that $\mathrm{Tr}\,\rho^2 \leq 1$, this gives us the desired bound:
\beq
\sum_{\vec{\alpha}} |W_{\vec{\alpha}}| \leq D \sqrt{\sum_{\vec{\alpha}} W_{\vec{\alpha}}^2} = \sqrt{D} \sqrt{\mathrm{Tr}\,\rho^2}  \leq  \sqrt{D}.
\eeq

\subsection{Quantum uncertainty and coherence}
\begin{figure}[h!]
\centering
\includegraphics[height=6cm]
{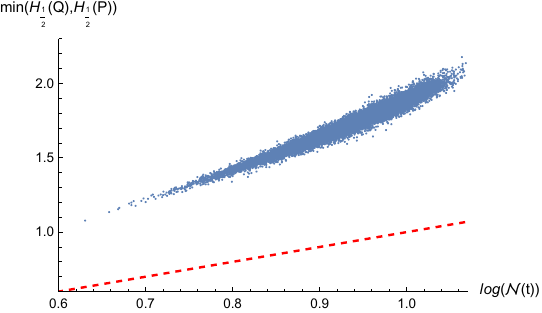}
\caption{A plot of $\min(H_{1/2}(q),H_{1/2}(p)$ vs log negativity for 30000 random pure states at $D=13$. The red dotted line indicates the lower bound in equation \eqref{inequality}. }
\label{fig:uncertainty}
\end{figure}

A second connection between Wigner negativity and non-classicality involves quantum uncertainty. One way to quantify quantum uncertainty is in terms of the ``spread'' of the wavefunction in position and momentum basis. For instance, a standard measure that is often used is the Shannon entropy:
\beq 
H_{\psi}(Q) = -\sum_q \lambda_q\,\ln \lambda_q,\;\; \lambda_q = |\langle q|\psi\rangle|^2,
\eeq 
\beq 
H_{\psi}(P) = -\sum_p \lambda_p\,\ln \lambda_p,\;\; \lambda_p = |\langle p|\psi\rangle|^2.
\eeq 
The Shannon entropies above are measures of the quantum spread of the wavefunction in phase space. There is an analog of the Heisenberg uncertainty principle in terms of these entropic measures which states that for any pure state $\psi$,
\beq 
H_{\psi}(P) + H_{\psi}(Q) \geq \log D,
\eeq 
where the inequality is saturated by stabilizer (i.e., Wigner positive) states. It is natural to ask how Wigner negativity is related to the above entropic measures of spreading in phase space. To this end, it is easy to prove the following simple bound (see also figure \ref{fig:uncertainty}):
\beq 
\text{min}\left(H^{(1/2)}_{\psi}(P), H^{(1/2)}_{\psi}(Q)\right) \geq \log \mathcal{N}(\psi),
\eeq 
where $H^{(1/2)}$ is the R\'enyi entropy:
\beq 
H_{\psi}^{(1/2)}(Q) = 2 \log\left(\sum_q \lambda_q^{1/2}\right).
\eeq 
In recent work, the Shannon entropy and the first moment of the above distributions have been studied as measures of the spreading of a quantum state under time evolution; these measures are commonly called spread complexity or Krylov complexity \cite{Balasubramanian:2022tpr}. The R\'enyi entropies for R\'enyi index less than one are also good measures of the spread of a probability distribution, and the above inequality gives a precise relation between Wigner negativity and spread complexity -- a Wigner negative state must unavoidably have some minimal spread, crucially in \emph{both} position and momentum. In fact, more generally, it is easy to show that
\beq \label{cliff}
\text{min}_{U\in \text{Cliff}}\,H^{(1/2)}_{\psi}(U\cdot Q) \geq \log \mathcal{N}(\psi),
\eeq 
where the minimization is over all Clifford unitaries. In other words, for a Wigner negative state, the wavefunction in any basis obtained from a Clifford rotation of the position basis must have some minimal spread. Clifford unitaries correspond to symplectic rotations \cite{Gross_2006}, so we can interpret equation \eqref{cliff} as the statement that a Wigner negative state must have a minimal spread in ``every direction'' in phase space. Note that the above bound is saturated by Wigner positive states; in this case the minimal spread $\text{min}_{U\in \text{Cliff}}\,H^{(1/2)}_{\psi}(U\cdot Q)$ also vanishes. In fact, the minimal spread for a pure state is zero if and only if the state is Wigner positive. 

The bound in equation \eqref{cliff} can be generalized to mixed states as well. In this case, the quantity that naturally appears on the left hand side is a R\'enyi version of the \emph{diagonal entropy}. Given a density matrix $\rho$ and an orthonormal basis $\mathcal{B}$, the diagonal entropy is defined as
\beq 
H_{\rho}(\mathcal{B}) = -\sum_i \lambda_i\log\lambda_i,\;\;\lambda_i = \langle i_{\mathcal{B}}|\rho|i_{\mathcal{B}}\rangle.
\eeq
The diagonal entropy \cite{Streltsov_2017} (see also \cite{Chandra:2022fwi} for a recent discussion in the gravitational context) plays an important role as a measure of \emph{coherence} in the resource theory of quantum coherence \cite{Streltsov_2017}.\footnote{We thank Pawel Caputa for helpful discussions on this point.}  For instance, it is easy to show that the relative entropy of coherence, defined as the minimum possible value of the relative entropy between $\rho$ and any incoherent state (i.e., a state which is diagonal with respect to $\mathcal{B})$, is given by: 
$$C(\rho,\mathcal{B}) := H_{\rho}(\mathcal{B}) - S_{\text{vN}}(\rho),$$
where $S_{\text{vN}}(\rho)$ is the von Neumann entropy of $\rho$. In the resource theory of quantum coherence, incoherent states are regarded as classical, and measures such as the relative entropy of coherence quantify the non-classicality of a state. In our case, the generalization of the inequality \eqref{cliff} to mixed states involves a R\'enyi version of the diagonal entropy:
\beq 
H_{\rho}^{(1/2)}(\mathcal{B}) = 2\log\left(\sum_i \lambda^{1/2}_i\right),\;\;\lambda_i = \langle i_{\mathcal{B}}|\rho|i_{\mathcal{B}}\rangle.
\eeq 
This gives yet another connection between two interesting measures of non-classicality, namely coherence and magic. 

We now turn to the proof of the above inequalities. Recall that the Wigner negativity for a general (possibly mixed) state $\rho$ is given by:
\beqn
 \cN(\rho) = \sum_{q,p} \left| W_{\rho}(q,p)\right|,\;\;W_{\rho}(q,p)=\frac{1}{D} \sum_{k,\ell=0}^{D-1} \widehat{\delta}_{2q,k+\ell} e^{\frac{2\pi i}{D}(k-\ell)p} \langle k|\rho |\ell\rangle .
\eeqn

Using the triangle inequality, we can write:
\beqn
 \cN(\rho) &\leq& \sum_{q,p} \frac{1}{D} \sum_{k,\ell=0}^{D-1} \widehat{\delta}_{2q,k+\ell} \left| \langle k|\rho |\ell\rangle \right|
 \nonumber \\
& \leq & \sum_{q,p} \frac{1}{D} \sum_{k,\ell=0}^{D-1} \widehat{\delta}_{2q,k+\ell}  \sqrt{\langle k|\rho |k\rangle}\sqrt{\langle \ell|\rho |\ell\rangle}
 \nonumber \\
 &=& \left( \sum_{k=0}^{D-1} \sqrt{\langle k|\rho |k\rangle} \right)^2,
 \nonumber \\
 \implies \log(\cN(\rho)) &\leq& 2 \log \left( \sum_{q=0}^{D-1} \sqrt{\langle k|\rho |k\rangle} \right) = H^{(1/2)}_{\rho}(Q).
 \nonumber
\eeqn
In the second line, we have used the Cauchy-Schwarz inequality. Since the Wigner function is symmetric with respect to position and momentum basis (i.e., we would have obtained the same  Wigner function had we replaced the position basis with momentum basis in our formulas), we must also have:
\beqn
\log(\cN(\rho)) \leq H^{(1/2)}_{\rho}(P),
\eeqn
and thus
\beqn
\log(\cN(\rho)) \leq \min \left( H^{(1/2)}_{\rho}(Q), H^{(1/2)}_{\rho}(P) \right).
\label{inequality}
\eeqn
More generally, the Wigner function transforms covariantly under any Clifford unitary, and so the negativity remains invariant under such Clifford unitary transformations. This immediately implies equation \eqref{cliff}. 

We have already discussed several ways in which Wigner negativity serves as a measure of non-classicality, or inherent ``quantum-ness''. There is another way to understand this connection for multi-partite systems from the point of view of ``contextuality'' \cite{Howard:2014zwm, Delfosse_2017}. Wigner positive states admit a non-contextual, hidden variable model description. Furthermore, it can be shown that Wigner negativity is an obstruction to the existence of such a description \cite{Howard:2014zwm, Delfosse_2017}. This provides yet another perspective on why Wigner negativity should be thought of as a measure of non-classicality. We refer the interested reader to the above references for further details.

\subsection{Minimizing negativity growth}
We now come back to the question that we started off with: when does the quantum time evolution of a state admit a semi-classical description? Given the interpretation of Wigner negativity as a measure of non-classicality, one might say that this happens to be the case if the Wigner negativity of the time-evolved state remains ``as small as possible'' at all times. However, recall that the Wigner function is defined with respect to a choice of an ordered, orthonormal basis. Thus, it seems natural to formulate the above question as follows: given an initial state $\psi_0$ and a time evolution operator $e^{-itH}$, can we find an ordered, orthonormal basis with respect to which the growth of Wigner negativity under time evolution is minimized? If so, one might regard the dynamics in this basis as an \emph{effective classical description} of the quantum dynamics.

There are several things which need to be refined in the above formulation: firstly, we do not need a basis for the entire Hilbert space, but only for the subspace spanned by the time evolution of the initial state. However, for the most part, we will be interested in the time evolution of generic initial states under chaotic Hamiltonians; in this setting, we expect the time evolution to be ergodic, i.e., that it should span the entire Hilbert space. The criterion on the initial state being generic is necessary here, because for special states, say for e.g., energy eigenstates, the time evolution becomes trivial multiplication by a phase. Secondly, we need to quantify what we mean by the negativity remaining small under time evolution -- our criterion will be that for any finite, $O(1)$ amount of time evolution, the negativity should remain $O(1)$ and not start scaling with the Hilbert space dimension.   

In \cite{2024JHEP...05..264B}, it was proposed that the Krylov basis (up to multiplication by individual phases) minimizes the early time Wigner negativity growth in the large $D$ limit. The Krylov basis is defined as follows: start with the set of states $\psi_0, H\psi_0, H^2\psi_0, \cdots$, and construct an orthonormal set of states by following Gram-Schmidt orthogonalization. The argument given in \cite{2024JHEP...05..264B} was perturbative in time: if we wish to minimize the negativity at $t=0$, we can simply take $\psi_0$ to be a basis vector, and without loss of generality, we take it to be the $0$th basis vector. Then, it was shown that for any choice of the first basis vector in the putative basis, the coefficient of the linear in time growth of the negativity is always larger than the coefficient of the linear in time growth in the Krylov basis. Similarly, for any basis which agrees with the Krylov basis up to the $m$th vector, but differs at index $m+1$, the coefficients in the Taylor approximation of the Wigner function agree between the two bases up to $t^m$, but at $O(t^{m+1})$, the negativity in the Krylov basis is smaller than any other basis. In \cite{2024JHEP...05..264B}, it was shown that this perturbative reasoning forces one to pick the Krylov basis up to individual phases, where the phases were left undetermined. 

The above argument strongly suggests that the Krylov basis minimizes Wigner negativity growth. However, one significant drawback with this argument is that it is perturbative in time. In this paper, our main goal is to establish results about the finite time evolution of Wigner negativity in the Krylov basis, and in particular, compare this with other choices of bases. We will prove two main results: for a generic choice of basis vectors (i.e., chosen at random, without any fine tuning), the negativity grows rapidly to an exponentially large value (more precisely, starts scaling with $D^{1/2}$) within an $O(1)$ amount of time evolution. On the other hand, in the Krylov basis, at any large but $O(1)$ time, the Wigner negativity grows at most as a power law in time, and thus does not scale with $D$ for any sub-exponential time. In proving these results, we will resort to tools from random matrix theory. For specific choices of Hamiltonian, the time evolution of the Wigner negativity is hard to compute explicitly. However, as we will show, the average behavior of negativity -- where the average is over a suitable ensemble of Hamiltonians -- is more tractable. In particular, we will establish both the above results in the context of Hamiltonians chosen randomly from the Gaussian unitary ensemble. Since chaotic Hamiltonians are, on general grounds, expected to satisfy random matrix theory statistics, we expect that the qualitative features of our results will be valid more generally for all chaotic Hamiltonians.

\section{Wigner negativity growth in a generic basis}\label{sec:generic}
In this section, we will study the time evolution of Wigner negativity with respect to a generic choice of basis. 
\subsection{Ensemble average and replica trick}
  We will focus on the case of chaotic Hamiltonians with no (local) conserved charges. The time evolution of the Wigner negativity for one specific Hamiltonian is, in general, difficult to compute. In order to make progress, we will need to resort to tools from random matrix theory. We will take the Hamiltonian to be a random matrix with respect to some basis, i.e., we assume that for some choice of basis which we will call the \emph{computational basis} $\mathcal{B}=\{|i_0\rangle\}_{i=0}^{D-1}$, the matrix elements $H_{ij}$ of the Hamiltonian in this basis constitute a random matrix chosen from some unitarily invariant ensemble:
\beq 
\mu(H) = \frac{1}{Z}\,e^{-D\,\mathrm{Tr}\,V(H)}.
\eeq 
We will use the computational basis to define the Wigner function, and take the initial state $|\psi_0\rangle$ to be the first vector $|0\rangle$ of the computational basis. Since the Hamiltonian looks like a random matrix in this basis, the computational basis is a generic choice of basis and the vector $|0\rangle$ is a generic initial vector as far as time evolution is concerned. With this choice, the Wigner function is everywhere non-negative at $t=0$, i.e., $\mathcal{N}(\psi_0)=1$. In fact, we could have taken any pure classical (i.e., Wigner positive) state as our initial state. Since every classical state can be obtained from $|0\rangle$ by a Clifford unitary, the ensemble-averaged negativity at all times is unaffected. Our aim is to calculate the time evolution of the negativity. Of course, the calculation for any one choice of Hamiltonian is hard, so we will instead calculate the ensemble average of the Wigner negativity:
\beq 
\overline{\mathcal{N}(\psi_t)} = \int dH\,\mu(H) \,\mathcal{N}(\psi_t),\;\;\;\psi_t = e^{-itH}\psi_0.
\eeq 
It will turn out that the ensemble average of the negativity is not the same as the negativity of the ensemble averaged Wigner function, even for $O(1)$ times. 

In order to compute the averaged negativity, we will use the following trick: the Wigner function in the computational basis is given by:
\beq 
\;W_{\psi,\mathcal{B}}(q,p)=\frac{1}{D} \sum_{k,\ell=0}^{D-1} \widehat{\delta}_{2q,k+\ell} e^{\frac{2\pi i}{D}(k-\ell)p} \langle k_0|e^{-itH}|\psi_0\rangle \langle \psi_0|e^{itH} |\ell_0\rangle,
\eeq 
where the subscript $\mathcal{B}$ is meant to indicate that the Wigner function is defined with respect to the basis $\mathcal{B}$. Now suppose we consider the Wigner function in a different basis $\mathcal{B}'=\{|i\rangle\}_{i=0}^{D-1}$ given by a unitary rotation $U$ of $\mathcal{B}$, but such that the unitary leaves the state $\psi_0$ invariant:
\beq 
|i\rangle = \begin{cases}|\psi_0\rangle\;\;\; \cdots\;\;\; i=0\\\sum_{j=1}^{D-1}U_{ij}|j_0\rangle\;\;\; \cdots\;\;\; i\neq 0,\end{cases} \label{eq:new_basis}
\eeq
or equivalently
\beq
|i\rangle = \delta_{i,0}|\psi_0\rangle + \bdelta_{i,0}\,\sum_{j=1}^{D-1}U_{ij}|j_0\rangle, \label{eq:new_basis_2}
\eeq
where $U_{ij} = \langle j_0| U |i_0\rangle$, and we have defined $\bdelta_{m,n} = (1-\delta_{m,n})$. The Wigner function in the new basis is given by
\beqn 
\;W_{\psi,U\cdot\mathcal{B}}(q,p)&=& \frac{1}{D} \sum_{k,\ell=0}^{D-1} \widehat{\delta}_{2q,k+\ell} e^{\frac{2\pi i}{D}(k-\ell)p} \langle k_0| U^{\dagger}\,e^{-itH}|\psi_0\rangle \langle \psi_0|e^{itH} U|\ell_0\rangle\nonumber\\
&=& \frac{1}{D} \sum_{k,\ell=0}^{D-1} \widehat{\delta}_{2q,k+\ell} e^{\frac{2\pi i}{D}(k-\ell)p} \langle k_0| U^{\dagger}\,e^{-itH}U|\psi_0\rangle \langle \psi_0|U^{\dagger}e^{itH} U|\ell_0\rangle,
\eeqn 
where in the second line, we have used the fact that $U|\psi_0\rangle = |\psi_0\rangle$. But inserting the Wigner function in the averaged negativity, we see that the negativity:
\beq 
\overline{\mathcal{N}_{U.\mathcal{B}}(\psi_t)} = \int dH\,\mu(H) \sum_{q,p}\left| W_{\psi_t,U.\mathcal{B}}(q,p)\right|,
\eeq 
is independent of $U$; this follows from the unitary invariance of the ensemble, i.e., the fact that the integration measure $dH \mu(H)$ is invariant under the redefinition $H' = U^{\dagger}HU$. Thus, 
\beq
\overline{\mathcal{N}_{\mathcal{B}}(\psi_t)} =\overline{\mathcal{N}_{U\cdot\mathcal{B}}(\psi_t)}.
\eeq 
Since the averaged negativity is invariant under such unitary transformations, we can integrate the averaged negativity over $U$ with the Haar measure: 
\beq 
\overline{\mathcal{N}_{\mathcal{B}}(\psi_t)} =\frac{1}{\text{Vol}_{U(D-1)}}\int dU\int dH\,\mu(H) \sum_{q,p}\left| W_{\psi_t,U.\mathcal{B}}(q,p)\right|,
\eeq 
where the Haar integration is over the group $U(D-1)$. So far, all we have done is introduced a fictitious integral over $U$ which has no effect. We will now switch the order of the two integrations above, i.e., move the $U$ integral inside the $H$ integral:
\beq \label{HAve}
\overline{\mathcal{N}_{\mathcal{B}}(\psi_t)} =\frac{1}{\text{Vol}_{U(D-1)}}\int dH\,\mu(H) \int dU\sum_{q,p}\left| W_{\psi_t,U.\mathcal{B}}(q,p)\right|.
\eeq 
As we will see below, we will now be able to carry out the integration over $U$ by using standard methods for Haar integration. The essence of the above manipulations is the simple point that the integration measure over $H$ can be split up as into an integral over the energy eigenvalues times an integral over the unitary group $U(D)$: 
\beq 
dH \mu(H) = d\mathcal{U}\, d^D\lambda\,\Delta(\vec{\lambda}) \mu(\vec{\lambda}).
\eeq 
The $U$ integral in equation \eqref{HAve} can be thought of as coming from the Haar integration above. It is not quite the full integral over the unitary group $U(D)$, but only a part of it over the subgroup $U(D-1)$ which leaves $\psi_0$ invariant. The reason we have chosen to present the calculation in this way is that the integral:
\beq \label{IntI}
I = \frac{1}{\text{Vol}_{U(D-1)}}\int dU\sum_{q,p}\left| W_{\psi_t,U.\mathcal{B}}(q,p)\right|
\eeq 
has the interpretation as the negativity of the Wigner function with respect to a basis which has $\psi_0$ as its first vector, but where the remaining basis vectors are chosen randomly. From this point of view, the Hamiltonian is fixed but the randomness comes from picking a generic basis.

Our goal now is to compute the integral $I$ in equation \eqref{IntI}. For any polynomial in $U$ and $U^{\dagger}$, the integral over $U$ can be performed using the standard theory of Haar integration. However, one problem in our case is that we need to perform the integral over the absolute value of the Wigner function, which is not an analytic function in $U$. In order to overcome this difficulty, we will resort to the \emph{replica trick} (see also \cite{Leone2} for work on R\'enyi versions of Wigner negativity). Instead of computing the integral at hand, we instead compute the following integral:
\beq 
I_{n} = \frac{1}{\text{Vol}_{U(D-1)}}\int dU\,\sum_{q,p}W^{2n}_{\psi_t,U\cdot \mathcal{B}}(q,p),
\eeq
for some positive integer $n$. Note that $I_{n}$ involves products of $U$ and $U^{\dagger}$, so this integral can be evaluated using standard techniques from the theory of Haar integration. If the final answer is analytic in $n$, then we can obtain the original integral $I$ by analytically continuing $I_n$ to $n=\frac{1}{2}$.\footnote{While the analytic continuation trick is a bit ad hoc, it is good enough for our purposes and gives results consistent with numerical calculations. A more systematic approach without resorting to analytic continuation is also possible, and is in fact necessary while calculating the evolution of Wigner negativity for open quantum systems, such as evaporating black holes \cite{upcoming}.}

\subsection{Diagrammatic Method}\label{rulesdiags}
The Wigner function at time $t$ in the basis $U\cdot \mathcal{B}$ is given by
\beq 
W_{\psi(t)}(q,p) = \frac{1}{D}\sum_{i,i'=0}^{D-1}e^{\frac{2\pi i p}{D}(i-i')} \widehat{\delta}_{2q,i+i'}\left(\delta_{i,0}\ s + \sum_{j=1}^{D-1}\bdelta_{i,0}U_{ij}s_j\right) \left(\delta_{i',0}\ \bar{s} + \sum_{j=1}^{D-1}\bdelta_{i',0}U^*_{i' j'}\bar{s}_{j'}\right),
\label{wignerfunction}
\eeq
where $s = \langle \psi(t)|\psi_0\rangle$ and $s_{j} = \langle \psi(t)|j_0\rangle$. We will sometimes also use the notation 
$$ \bar{s}s = S(t),\;\;\sum_{j=1}^{D-1}\bar{s_{j}}s_{j} = 1-S(t),$$ 
where $S(t) = |\langle \psi_0|e^{-itH}|\psi_0\rangle|^2$ is the survival probability. 
Our goal now is to compute the Haar averaged Réyni Wigner negativity:
\begin{equation}
\begin{split}
I_n = \int \frac{dU}{\text{Vol}_{U(D-1)}}  \sum_{q,p}W_{\psi_t}^{2n},
\label{Reyninegativity}
\end{split}
\end{equation}
\beq
W_{\psi_t}^{2n} = \prod_{k=1}^{2n}\left[\frac{1}{D}\sum_{i_k,i_k'=0}^{D-1}e^{\frac{2\pi i p}{D}(i_k-i_k')} \widehat{\delta}_{2q,i_k+i_k'}\left(\delta_{i_k,0}\ s + \sum_{j_k=1}^{D-1}\bdelta_{i_k,0}U_{i_kj_k}s_{j_k}\right) \left(\delta_{i_k',0}\ \bar{s} + \sum_{j'_k=1}^{D-1}\bdelta_{i_k',0}U^*_{i_k' j_k'}\bar{s}_{j_k'}\right)\right].
\eeq 
The Haar average can be computed using the following general formula \cite{Mele2024introductiontohaar}:
\beq
\int  \frac{dU}{\text{Vol}_{U(D-1)}}\, U_{i_1j_1}\hdots 
 U_{i_pj_p}U^*_{i_1'j_1'}\hdots U^*_{i_p'j_p'}=
\sum_{\sigma,\tau\in S_p}
\delta_{i_1i_{\sigma(1)}'}\hdots\delta_{i_pi_{\sigma(p
)}'}\delta_{j_1 j_{\tau(1)}'}\hdots\delta_{j_pj_{\tau(p)}'}\operatorname{Wg}(\sigma\tau^{-1},D-1), 
\label{eq:Haar}
\eeq
where the functions $\operatorname{Wg}$ are called Weingarten functions.

We will find it convenient to use diagrammatic notation to perform this integral. We label each copy of a Wigner function with a blob containing an empty dot and a shaded dot -- the empty dot denotes the un-primed index $i_k$ (associated to the bra $\langle \psi_t|$) while the shaded dot denotes the primed index $i'_k$ (associated to the ket $| \psi_t\rangle$) in the above expression. Corresponding to the $2n$ powers of the Wigner function in the R\'enyi negativity, there will be $2n$ such blobs in any diagram. Each empty dot can either have no leg attached to it, corresponding to the $\delta_{i_k,0}$ term, or it can have one leg coming out of it, corresponding to the $\overline{\delta}_{i_k,0}$ term. Similarly, each shaded dot can either have no leg attached to it, corresponding to the $\delta_{i'_k,0}$ term, or it can have one leg going into it, corresponding to the $\overline{\delta}_{i'_k,0}$ term. There are of course the appropriate factors of $s$, $s_{j_k}$ etc. that must be kept track of. Then, equation \eqref{eq:Haar} implies that any leg coming out of an empty dot must be connected or contracted with a leg going into a shaded dot. There are also, of course, the $j$ and $j'$ indices, but we do not need to keep track of these indices in our notation -- the reason for this is that no matter how the j indices are contracted among each other, they always lead to an overall factor of $(1-S(t))^p$, where $p$ is the number of of contractions (i.e., lines joining empty and shaded dots). More explicitly, in any diagram with $p$ contractions, the Haar integral takes the form:
 \beq
 \begin{split}
 \int  \frac{dU}{\text{Vol}_{U(D-1)}}\left(\prod_{k=1}^ps_{j_k} U_{i_kj_k}\right)\left(\prod_{\ell=1}^p s_{j'_\ell}U^*_{i_\ell'j_\ell'}\right) &= (1- S(t))^p
\sum_{\sigma\in S_p}
\delta_{i_1i_{\sigma(1)}'}\cdots\delta_{i_pi_{\sigma(p
)}'}\sum_{\tau\in S_p}\operatorname{Wg}(\sigma\tau^{-1},D-1)\\
&=(1- S(t))^p\left(\sum_{\rho\in S_p}\operatorname{Wg}(\rho,D-1)\right)\sum_{\sigma\in S_p}
\delta_{i_1i_{\sigma(1)}'}\cdots\delta_{i_pi_{\sigma(p
)}'}\\
&=(1- S(t))^p\frac{(D-2)!}{(D-2+p)!}\sum_{\sigma\in S_p}
\delta_{i_1i_{\sigma(1)}'}\cdots\delta_{i_pi_{\sigma(p
)}'}.
\end{split}
\label{eq:Haarwithoutj}
 \eeq
Thus, from a diagrammatic point of view, we must simply contract all the open legs via Wick contractions, keeping in mind that empty dots contract with shaded dots. Another point to note is that on account of these Wick contractions, all the momentum dependent phases in equation \eqref{Reyninegativity} cancel out.

\subsubsection*{Averaged Wigner function}
Let us consider the simple example of the Haar average of the Wigner function:
\beq
\int  \frac{dU}{\text{Vol}_{U(D-1)}} W_{\psi(t)}(q,p)=\frac{1}{D} \sum_{i_1,i_1'}e^{\frac{2\pi i p}{D}(i_1-i_1')} \widehat{\delta}_{2q,i_1+i_1'}\int  \frac{dU}{\text{Vol}_{U(D-1)}}\left(\delta_{i_1,0}\ s + \bdelta_{i_1,0}U_{i_1j_1}s_{j_1}\right) \left(\delta_{i_1',0}\ \bar{s} + \bdelta_{i_1',0}U^*_{i_1' j_1'}\bar{s}_{j_1'}\right).
\label{onepointwigner}
\eeq
In terms of our diagrammatic notation, we have the four possible terms in the intgrand:
$$\begin{matrix} 
\begin{tikzpicture}
\coordinate (P) at (0,0);

\draw (P) circle (10pt); 

\node[below] at (0,-10pt) {1};

\coordinate (P1) at (-0.15,0);
\coordinate (P2) at (0.15,0);

\draw[black] (P1) circle (1.7pt);
\filldraw (P2) circle (1.7pt);

\draw[-, thick] (P1) to (-0.7,0.7);
\end{tikzpicture} & \hspace{1cm}
\begin{tikzpicture}
\coordinate (P) at (0,0);

\draw (P) circle (10pt); 

\node[below] at (0,-10pt) {1};

\coordinate (P1) at (-0.15,0);
\coordinate (P2) at (0.15,0);

\draw[black] (P1) circle (1.7pt);
\filldraw (P2) circle (1.7pt);


\draw[-, thick] (P2) to (0.7,0.7);
\end{tikzpicture}&\hspace{1cm}
\begin{tikzpicture}
\coordinate (P) at (0,0);

\draw (P) circle (10pt); 

\node[below] at (0,-10pt) {1};

\coordinate (P1) at (-0.15,0);
\coordinate (P2) at (0.15,0);

\draw[black] (P1) circle (1.7pt);
\filldraw (P2) circle (1.7pt);

\draw[-, thick] (P1) to (-0.7,0.7);
\draw[-, thick] (P2) to (0.7,0.7);
\end{tikzpicture} \hspace{1cm}
\begin{tikzpicture}
\coordinate (P) at (0,0);

\draw (P) circle (10pt); 

\node[below] at (0,-10pt){1};

\coordinate (P1) at (-0.15,0);
\coordinate (P2) at (0.15,0);

\draw[black] (P1) circle (1.7pt);
\filldraw (P2) circle (1.7pt);
\end{tikzpicture}
\end{matrix}$$
which represent the $\bar\delta_{i_1,0}\delta_{i_1^\prime,0}$; $\delta_{i_1,0}\bar\delta_{i_1^\prime,0}$; $\bar\delta_{i_1,0}\bar\delta_{i_1^\prime,0}$ and $\delta_{i_1,0}\delta_{i_1^\prime,0}$ contributions respectively. Performing the Haar average of the Wigner function will give Wick contractions of the limbs, as explained above. Thus, the first two configurations above vanish because there is no way to contract the open legs in these diagrams, and we are left with the following two contributions to the Haar integration:
$$\begin{matrix}
\begin{tikzpicture}
\coordinate (P) at (0,0);

\draw (P) circle (10pt);

\node[below] at (0,-10pt) {1};

\coordinate (P1) at (-4pt,0);
\coordinate (P2) at (4pt,0);

\draw[black] (P1) circle (1.7pt);
\filldraw (P2) circle (1.7pt);

\draw[black,out = 120,in = 60, min distance = 2cm] (P1) to (P2);
\end{tikzpicture}&
\begin{tikzpicture}
\coordinate (P) at (0,0);

\draw (P) circle (10pt); 

\node[below] at (0,-10pt){1};

\coordinate (P1) at (-0.15,0);
\coordinate (P2) at (0.15,0);

\draw[black] (P1) circle (1.7pt);
\filldraw (P2) circle (1.7pt);
\end{tikzpicture}
\end{matrix}$$
The first of these gives 
\beq 
\frac{1}{D(D-1)}(1 - S(t))\sum_{i_1,i_1'}e^{\frac{2\pi ip}{D}(i_1-i'_1)}\widehat{\delta}_{2q,i_1+i_1'} \delta_{i_1,i_1'}\overline{\delta}_{i_1,0}\overline{\delta}_{i'_1,0}= \frac{1}{D(D-1)}(1-S(t))\overline{\delta}_{q,0},
\eeq
while the second diagram gives 
\beq 
\frac{1}{D}S(t)\,\delta_{q,0}.
\eeq
Thus, the ensemble averaged Wigner function is given by
\beq
\overline{W_{\psi_t}(q,p)} =\frac{1}{D}\overline{S(t)}\,\delta_{q,0}+ \frac{1}{D(D-1)}(1-\overline{S(t)})\,\overline{\delta}_{q,0},
\eeq 
where $\overline{S}$ is the ensemble-averaged survival probability. Since $0\leq S(t) \leq 1$, we see that the averaged Wigner function is everywhere non-negative. One might be tempted to conclude from this that the Wigner negativity vanishes, but this is of course, incorrect -- the averaged negativity is not the same as the negativity of the averaged Wigner function. Note also that the averaged Wigner function is properly normalized, i.e., $\sum_{q,p}\overline{W_{\psi_t}(q,p)} = 1$. 

We can give an explicit formula for the averaged survival probability in terms of the spectral form factor. Recall that
 \beq
 \overline{S(t)}= \int dH\,\mu(H) \langle \psi_0|e^{-itH}|\psi_0 \rangle  \langle \psi_0|e^{+itH}|\psi_0 \rangle.
 \eeq 
Writing $H = \mathcal{U}\,h\,\mathcal{U}^{\dagger}$ where $h$ is diagonal and $\mathcal{U}$ is a Haar random unitary (this time a general unitary matrix in $U(D)$), we get
\beq 
\overline{S(t)} = \int dh\,\widetilde{\mu}(h)\int D\mathcal{U}\, \langle \psi_0|\mathcal{U} e^{-ith}\mathcal{U}^{\dagger}|\psi_0 \rangle  \langle \psi_0|\mathcal{U}e^{+ith}\mathcal{U}^{\dagger}|\psi_0 \rangle.
\eeq 
where $\widetilde{\mu}(h)$ is the measure for integration over energy eigenvalues (this includes the Van der Monde determinant). Performing the Haar integral over $\mathcal{U}$, we get
\beq \label{SFF}
\overline{S(t)} = \frac{1}{D^2}\int dh\,\widetilde{\mu}(h)\,\left[\text{Tr}(e^{-ith})\text{Tr}(e^{+ith})+\cdots\right],
\eeq 
where $\cdots$ denote lower order terms in $\frac{1}{D}$, which become important at late times (i.e., exponential times). The first term above is precisely the spectral form factor (at infinite temperature). The leading contribution to this, of course, comes from the disconnected answer (i.e., the product of the averages). For example, in GUE, this is given by 
\beq \label{SFFGUE}
\overline{S(t)} = \left(\frac{J_1(2t)}{t}\right)^2+ \cdots.
\eeq
This formula shows that the averaged-survival probability decays away from one as a function of time. It is known that the sub-leading corrections to this formula become important at a time of order $t\sim O(\sqrt{D})$ \cite{Cotler:2016fpe}.
\subsection{Evaluation of Wigner-Negativity}
We now turn to the calculation of the averaged Wigner negativity. Let us first begin by discussing some basic/elementary diagrams which will play an important role because they can appear as sub-diagrams in bigger diagrams. Firstly, there is the trivial diagram where we have a blob with no limbs attached. This diagram can be multiplied $k$ times to get a completely disconnected diagram with $k$ blobs. 
Say, for $k=3$ we have:\\
\newline
$$\begin{matrix}
\begin{tikzpicture}
\coordinate (P) at (0,0);
\coordinate (Q) at (40pt,0);
\coordinate (R) at (80pt,0);

\draw (P) circle (10pt);
\draw (Q) circle (10pt); 
\draw (R) circle (10pt); 

\node[below] at (0,-10pt) {1};
\node[below] at (40pt,-10pt) {2};
\node[below] at (80pt,-10pt) {3};

\coordinate (P1) at (-4pt,0);
\coordinate (P2) at (4pt,0);
\coordinate (Q1) at (36pt,0);
\coordinate (Q2) at (44pt,0);
\coordinate (R1) at (76pt,0);
\coordinate (R2) at (84pt,0);

\draw[black] (P1) circle (1.7pt);
\filldraw (P2) circle (1.7pt);
\draw[black] (Q1) circle (1.7pt);
\filldraw (Q2) circle (1.7pt);
\draw[black] (R1) circle (1.7pt);
\filldraw (R2) circle (1.7pt);


\end{tikzpicture}
\end{matrix}$$\\
Each of these disconnected diagrams individually represents a contribution of the type: $\delta_{i_m,0}\delta_{i_m^\prime,0}$ for each point $m$. So, the corresponding $\widehat{\delta}_{2q,i_m+i'_m}$ term in the Wigner function gives $\delta_{q,0}$. Thus, any diagram which has even one disconnected blob with no legs attached will be proportional to $\delta_{q,0}$, and gives a contribution that is localized at $q=0$. 

The other important class of fundamental diagrams are what we might call the ``fully contracted'' diagrams, where every blob comes with two legs attached, and these must all be contracted appropriately. Among the fully contracted diagrams, the connected ones are the basic building blocks. For each value of $k$, there is topologically only one such fully contracted, connected diagram; of course, one must sum over various permutations leading to the same topology. For instance, the $k=1,2,3$ fully contracted, connected diagrams are given by:\\
$$\begin{matrix}
\begin{tikzpicture}
\coordinate (P) at (0,0);

\draw (P) circle (10pt);

\node[below] at (0,-10pt) {1};

\coordinate (P1) at (-4pt,0);
\coordinate (P2) at (4pt,0);

\draw[black] (P1) circle (1.7pt);
\filldraw (P2) circle (1.7pt);

\draw[black,out = 120,in = 60, min distance = 2cm] (P1) to (P2);
\end{tikzpicture}&
\begin{tikzpicture}
\coordinate (P) at (0,0);
\coordinate (Q) at (40pt,0);

\draw (P) circle (10pt);
\draw (Q) circle (10pt); 

\node[below] at (0,-10pt) {1};
\node[below] at (40pt,-10pt) {2};

\coordinate (P1) at (0,4pt);
\coordinate (P2) at (0,-4pt);
\coordinate (Q1) at (40pt,-4pt);
\coordinate (Q2) at (40pt,4pt);

\draw[black] (P1) circle (1.7pt);
\filldraw (P2) circle (1.7pt);
\draw[black] (Q1) circle (1.7pt);
\filldraw (Q2) circle (1.7pt);

\draw[black,out = 60,in = 120, looseness = 1.5] (P1) to (Q2);
\draw[black,out = -60,in = -120, looseness = 1.5] (P2) to (Q1);
\end{tikzpicture}&
\hspace{5mm}
\begin{tikzpicture}
\coordinate (P) at (0,0);
\coordinate (Q) at (40pt,0);
\coordinate (R) at (80pt,0);

\draw (P) circle (10pt);
\draw (Q) circle (10pt); 
\draw (R) circle (10pt); 

\node[below] at (0,-10pt) {1};
\node[below] at (40pt,-10pt) {2};
\node[below] at (80pt,-10pt) {3};

\coordinate (P1) at (-4pt,0);
\coordinate (P2) at (4pt,0);
\coordinate (Q1) at (36pt,0);
\coordinate (Q2) at (44pt,0);
\coordinate (R1) at (76pt,0);
\coordinate (R2) at (84pt,0);

\draw[black] (P1) circle (1.7pt);
\filldraw (P2) circle (1.7pt);
\draw[black] (Q1) circle (1.7pt);
\filldraw (Q2) circle (1.7pt);
\draw[black] (R1) circle (1.7pt);
\filldraw (R2) circle (1.7pt);

\draw[black,out = 60,in = 120, looseness = 1.5] (P2) to (Q1);
\draw[black,out = 60,in = 120, looseness = 1.5] (Q2) to (R1);
\draw[black,out = 120,in = 60, looseness = 1.5] (R2) to (P1);
\end{tikzpicture}
\end{matrix}$$\\
 Let us now evaluate the fully contracted, connected diagrams with $k$ blobs. Up to an overall combinatorial factor corresponding to the number of permutations which give the fully connected topology with $k$ blobs, the contribution of such a diagram is given by 
\beq
C(k,q)=\sum_{i_1=1}^D\cdots\sum_{i_k=1}^D\widehat{\delta}_{2q,i_1+i_2}\widehat{\delta}_{2q,i_2 + i_3}\cdots\widehat{\delta}_{2q,i_{k-1}+i_k}\widehat{\delta}_{2q,i_k+i_1}.
\eeq
Let us first look at the case of $k$ even and $q=0$. In this case, we find $(D-1)$ solutions which are tabulated below:
\begin{table}[H]
    \centering
    \begin{tabular}{|c|c|c|c|c|}
    \hline
    $i_1$ & $i_2$ & $i_3 \hdots$ & $i_{k-1}$ & $i_k$ \\ \hline
    1     & $D-1$ & $1 \hdots$   & 1         & $D-1$ \\
    2     & $D-2$ & $2 \hdots$   & 2         & $D-2$ \\
    \vdots & \vdots & \vdots     & \vdots    & \vdots \\
    $D-1$ & 1     & $D-1 \hdots$ & $D-1$     & 1\\
    \hline
    \end{tabular}
\end{table}
Thus, for $k$ even and $q=0$, there are $(D-1)$ solutions, which gives $C(k=2n,0)=D-1$. On the other hand, when $k$ is odd, our $k$th column will no longer add up with our first column to give $D$ as it does here in the even case. This means that there are no solutions in this case, and thus $C(k=2n+1,0)=0$. Now consider the case where $q\neq 0$. Since no particular value of $q\neq 0$ is special, we can simply take $q=1$ without loss of generality. In the $k$ even case, the solutions are given by 
\begin{table}[H]
    \centering
    \begin{tabular}{|c|c|c|c|c|}
    \hline
    $i_1$ & $i_2$ & $i_3 \hdots$ & $i_{k-1}$ & $i_k$ \\ \hline
    1     & $1$ & $1\hdots$   &  $1$         & $1$   \\
    $3$     & $D-1$ & $3\hdots$   &  $3$         & $D-1$   \\
    \vdots & \vdots      & \vdots           & \vdots & \vdots \\
    $D-1$   & $3$ &$D-1\hdots$     & $ D-1$       & $3$   \\
    \hline
    \end{tabular}
\end{table}
Note that $i_1=2$ is not included here because that would correspond to $i_2=0$, which is not allowed. So, there are $D-2$ such cases.
Once again, we see that for the even case, the $k$th column and first column add up to give $(i_k+i_1)=2q$, just as we need. However, the only way this is possible for the odd $k$ case is when we take all the indices equal to $q$. In the specific example where $q=1$, this corresponds to the first row of the table above. Therefore, when $q\neq 0$, we have $C(k,q)=(D-2)$ for the even $k$ case and $C(k,q)=1$ for the odd $k$ case. So, to summarize, we get
\beq\label{ckq}
C(k,q) = \begin{cases} \overline{\delta}_{q,0} & \cdots (k=2n+1)\\ (D-1)\delta_{q,0}+(D-2)\overline{\delta}_{q,0} & \cdots (k=2n)\end{cases}
\eeq

From here on, we will work in the large $D$ limit. At large $D$, the leading contribution among the fully contracted diagrams in the Haar average of $W_{\psi_t}^{2n}$ comes from the diagram with pairwise contractions. This follows from our analysis above because the pairwise contracted diagram  has $n$ $k=2$ sub-diagrams, each of which we know gives $D$ -- so the pairwise contraction gives a factor of $D^n$. On the other hand, including a connected sub-diagram with three of more blobs naturally gives a lower power of $D$. For example, in the case of $W_{\psi_t}^4$, the pairwise contracted diagram is shown below and gives a factor of $D^2$, which is the leading contribution among all the fully contracted diagrams.
$$\begin{matrix}
\begin{tikzpicture}
\coordinate (P) at (0,0);
\coordinate (Q) at (40pt,0);

\draw (P) circle (10pt);
\draw (Q) circle (10pt); 

\node[below] at (0,-10pt) {1};
\node[below] at (40pt,-10pt) {2};

\coordinate (P1) at (0,4pt);
\coordinate (P2) at (0,-4pt);
\coordinate (Q1) at (40pt,-4pt);
\coordinate (Q2) at (40pt,4pt);

\draw[black] (P1) circle (1.7pt);
\filldraw (P2) circle (1.7pt);
\draw[black] (Q1) circle (1.7pt);
\filldraw (Q2) circle (1.7pt);

\draw[-,out = 60,in = 120, looseness = 1.5] (P1) to (Q2);
\draw[-,out = -60,in = -120, looseness = 1.5] (P2) to (Q1);
\end{tikzpicture}&
\begin{tikzpicture}
\coordinate (P) at (0,0);
\coordinate (Q) at (40pt,0);

\draw (P) circle (10pt);
\draw (Q) circle (10pt); 

\node[below] at (0,-10pt) {3};
\node[below] at (40pt,-10pt) {4};

\coordinate (P1) at (0,4pt);
\coordinate (P2) at (0,-4pt);
\coordinate (Q1) at (40pt,-4pt);
\coordinate (Q2) at (40pt,4pt);

\draw[black] (P1) circle (1.7pt);
\filldraw (P2) circle (1.7pt);
\draw[black] (Q1) circle (1.7pt);
\filldraw (Q2) circle (1.7pt);

\draw[-,out = 60,in = 120, looseness = 1.5] (P1) to (Q2);
\draw[-,out = -60,in = -120, looseness = 1.5] (P2) to (Q1);
\end{tikzpicture}
\end{matrix}$$

So far we have discussed two types of diagrams: the ones which include no limbs, and the ones where every blob has two limbs attached to it, with all the limbs contracted appropriately. The rest of the diagrams can be constructed by starting with a fully contracted diagram and then erasing legs one by one from the diagram. For example, if we remove one limb from the fully contracted, connected diagrams consisting of 2, 3 or 4 blobs, then the resulting diagrams look like:\\
$$\begin{matrix}
\begin{tikzpicture}
\coordinate (P) at (0,0);
\coordinate (Q) at (40pt,0);

\draw (P) circle (10pt);
\draw (Q) circle (10pt); 

\node[below] at (0,-10pt) {1};
\node[below] at (40pt,-10pt) {2};

\coordinate (P1) at (0,4pt);
\coordinate (P2) at (0,-4pt);
\coordinate (Q1) at (40pt,-4pt);
\coordinate (Q2) at (40pt,4pt);

\draw[black] (P1) circle (1.7pt);
\filldraw (P2) circle (1.7pt);
\draw[black] (Q1) circle (1.7pt);
\filldraw (Q2) circle (1.7pt);

\draw[solid,out = 60,in = 120, looseness = 1.5] (P1) to (Q2);
\end{tikzpicture}&&&&
\begin{tikzpicture}
\coordinate (P) at (0,0);
\coordinate (Q) at (40pt,0);
\coordinate (R) at (80pt,0);

\draw (P) circle (10pt);
\draw (Q) circle (10pt); 
\draw(R) circle (10pt);

\node[below] at (0,-10pt) {1};
\node[below] at (40pt,-10pt) {2};
\node[below] at (80pt,-10pt) {3};

\coordinate (P1) at (-4pt,0);
\coordinate (P2) at (4pt,0);
\coordinate (Q1) at (36pt,0);
\coordinate (Q2) at (44pt,0);
\coordinate (R1) at (76pt,0);
\coordinate (R2) at (84pt,0);

\draw[black] (P1) circle (1.7pt);
\filldraw (P2) circle (1.7pt);
\draw[black] (Q1) circle (1.7pt);
\filldraw (Q2) circle (1.7pt);
\draw[black] (R1) circle (1.7pt);
\filldraw (R2) circle (1.7pt);

\draw[-,out = 60,in = 120, looseness = 1.5] (P2) to (Q1);
\draw[-,out = 60,in = 120, looseness = 1.5] (Q2) to (R1);
\end{tikzpicture}&&&&
\begin{tikzpicture}
\coordinate (P) at (0,0);
\coordinate (Q) at (40pt,0);
\coordinate (R) at (80pt,0);
\coordinate (S) at (120pt,0);

\draw (P) circle (10pt);
\draw (Q) circle (10pt); 
\draw (R) circle (10pt);
\draw (S) circle (10pt); 

\node[below] at (0,-10pt) {1};
\node[below] at (40pt,-10pt) {2};
\node[below] at (80pt,-10pt) {3};
\node[below] at (120pt,-10pt) {4};

\coordinate (P1) at (-4pt,0);
\coordinate (P2) at (4pt,0);
\coordinate (Q1) at (36pt,0);
\coordinate (Q2) at (44pt,0);
\coordinate (R1) at (76pt,0);
\coordinate (R2) at (84pt,0);
\coordinate (S1) at (116pt,0);
\coordinate (S2) at (124pt,0);

\draw[black] (P1) circle (1.7pt);
\filldraw (P2) circle (1.7pt);
\draw[black] (Q1) circle (1.7pt);
\filldraw (Q2) circle (1.7pt);
\draw[black] (R1) circle (1.7pt);
\filldraw (R2) circle (1.7pt);
\draw[black] (S1) circle (1.7pt);
\filldraw (S2) circle (1.7pt);

\draw[-,out = 60,in = 120, looseness = 1.5] (P2) to (Q1);
\draw[-,out = 60,in = 120, looseness = 1.5] (R2) to (S1);
\draw[-,out = 60,in = 120, looseness = 1.5] (Q2) to (R1);
\end{tikzpicture}\\





\end{matrix}$$\\
respectively. After setting the indices with no legs attached to $0$, the $k$-blob diagram (up to an overall combinatorial factor) gives: 
$$\sum_{i_2,i_3,..i'_k=1}^{D-1}\hat{\delta}_{2q,i_2}\hat{\delta}_{2q,i_2 + i_3}\cdots\hat{\delta}_{2q,i_{k-1}+i_k}\hat{\delta}_{2q,i'_k}.$$
Note that all these contributions for $k \geq 3$ vanish. Consider, for instance, the $k=3$ case. The first Kronecker delta vanishes if $q=0$, so let us take $q\neq 0$. In this case, the first Kronecker delta sets $i_2 = 2q$. But then the second Kronecker delta vanishes, because $i_3=0$ is not included in the sum. A similar argument also applies to higher $k$. Therefore, only the $k=2$ case among the above diagrams survives, and evaluates to $\delta_{q,\overline{0}}$.

We are finally in a position to compute the Haar average of the R\'enyi Wigner negativity $I_n$.

\subsubsection*{Contribution from $q\neq 0$}
First, let us consider the contribution to the negativity coming from $q\neq 0$. We start from the fully contracted diagram (recall that pairwise Wick contractions dominate):
$$\begin{matrix}
\begin{tikzpicture}
\coordinate (P) at (0,0);
\coordinate (Q) at (40pt,0);

\draw (P) circle (10pt);
\draw (Q) circle (10pt); 

\node[below] at (0,-10pt) {1};
\node[below] at (40pt,-10pt) {2};

\coordinate (P1) at (0,4pt);
\coordinate (P2) at (0,-4pt);
\coordinate (Q1) at (40pt,-4pt);
\coordinate (Q2) at (40pt,4pt);

\draw[black] (P1) circle (1.7pt);
\filldraw (P2) circle (1.7pt);
\draw[black] (Q1) circle (1.7pt);
\filldraw (Q2) circle (1.7pt);

\draw[-,out = 60,in = 120, looseness = 1.5] (P1) to (Q2);
\draw[-,out = -60,in = -120, looseness = 1.5] (P2) to (Q1);
\end{tikzpicture}\\
\begin{tikzpicture}
\coordinate (P) at (0,0);
\coordinate (Q) at (40pt,0);

\draw (P) circle (10pt);
\draw (Q) circle (10pt); 

\node[below] at (0,-10pt) {3};
\node[below] at (40pt,-10pt) {4};

\coordinate (P1) at (0,4pt);
\coordinate (P2) at (0,-4pt);
\coordinate (Q1) at (40pt,-4pt);
\coordinate (Q2) at (40pt,4pt);

\draw[black] (P1) circle (1.7pt);
\filldraw (P2) circle (1.7pt);
\draw[black] (Q1) circle (1.7pt);
\filldraw (Q2) circle (1.7pt);

\draw[-,out = 60,in = 120, looseness = 1.5] (P1) to (Q2);
\draw[-,out = -60,in = -120, looseness = 1.5] (P2) to (Q1);
\end{tikzpicture}\\
\vdots\\
\begin{tikzpicture}
\coordinate (P) at (0,0);
\coordinate (Q) at (40pt,0);

\draw (P) circle (10pt);
\draw (Q) circle (10pt); 

\node[below] at (0,-10pt) {2n-1};
\node[below] at (40pt,-10pt) {2n};

\coordinate (P1) at (0,4pt);
\coordinate (P2) at (0,-4pt);
\coordinate (Q1) at (40pt,-4pt);
\coordinate (Q2) at (40pt,4pt);

\draw[black] (P1) circle (1.7pt);
\filldraw (P2) circle (1.7pt);
\draw[black] (Q1) circle (1.7pt);
\filldraw (Q2) circle (1.7pt);

\draw[-,out = 60,in = 120, looseness = 1.5] (P1) to (Q2);
\draw[-,out = -60,in = -120, looseness = 1.5] (P2) to (Q1);
\end{tikzpicture}
\end{matrix}= \frac{(2n|2)}{D^{2n}} \times \frac{1}{D^{2n}} \times D^2 \times D^{n} \times (1-S(t))^{2n}. $$
The factor of $(2n|2) = \frac{(2n)!}{2^n n!}$ is a combinatorial factor and comes from the number of ways of pairing $2n$ blobs into $n$ pairs. The first factor of $\frac{1}{D^{2n}}$ comes from the $\frac{1}{D}$ factor in each copy of the Wigner function (see the definition in equation \eqref{wignerfunction}). The second factor of $\frac{1}{D^{2n}}$ comes from the sum over Weingarten functions in equation \eqref{eq:Haarwithoutj}, where we have only kept the leading behavior in $\frac{1}{D}$; roughly speaking, every line gives a $\frac{1}{D}$ suppression due to the Weingarten functions. The factor of $D^2$ comes from summing over $(q,p)$ (it should have been $D(D-1)$, but we only keep the leading dependence). The factor of $D^n$ comes because every fully contracted, connected diagram with even blobs gives a factor of $D$ (see equation \eqref{ckq} and the discussion below it). Finally, the factor of $(1-S)^{2n}$ comes from the sums over $s_{j'}s^*_{j'}$, where the rule is that every line comes with one factor of $(1-S)$. So overall, this diagram gives the contribution:
$$\frac{(2n|2)}{D^{3n-2}} (1-S(t))^{2n}.$$
Interestingly, there are more diagrams which contribute at the same order. To see this, consider the following diagram, where we erase the top leg from the first contracted pair in the above fully contracted diagram:
$$\begin{matrix}
\begin{tikzpicture}
\coordinate (P) at (0,0);
\coordinate (Q) at (40pt,0);

\draw (P) circle (10pt);
\draw (Q) circle (10pt); 

\node[below] at (0,-10pt) {1};
\node[below] at (40pt,-10pt) {2};

\coordinate (P1) at (0,4pt);
\coordinate (P2) at (0,-4pt);
\coordinate (Q1) at (40pt,-4pt);
\coordinate (Q2) at (40pt,4pt);

\draw[black] (P1) circle (1.7pt);
\filldraw (P2) circle (1.7pt);
\draw[black] (Q1) circle (1.7pt);
\filldraw (Q2) circle (1.7pt);

\draw[-,out = -60,in = -120, looseness = 1.5] (P2) to (Q1);
\end{tikzpicture}\\
\begin{tikzpicture}
\coordinate (P) at (0,0);
\coordinate (Q) at (40pt,0);

\draw (P) circle (10pt);
\draw (Q) circle (10pt); 

\node[below] at (0,-10pt) {3};
\node[below] at (40pt,-10pt) {4};

\coordinate (P1) at (0,4pt);
\coordinate (P2) at (0,-4pt);
\coordinate (Q1) at (40pt,-4pt);
\coordinate (Q2) at (40pt,4pt);

\draw[black] (P1) circle (1.7pt);
\filldraw (P2) circle (1.7pt);
\draw[black] (Q1) circle (1.7pt);
\filldraw (Q2) circle (1.7pt);

\draw[-,out = 60,in = 120, looseness = 1.5] (P1) to (Q2);
\draw[-,out = -60,in = -120, looseness = 1.5] (P2) to (Q1);
\end{tikzpicture}\\
\begin{tikzpicture}
\filldraw (0,0) circle (0.5pt);
\filldraw (0,4pt) circle (0.5pt);
\filldraw (0,8pt) circle (0.5pt);
\end{tikzpicture}\\
\begin{tikzpicture}
\coordinate (P) at (0,0);
\coordinate (Q) at (40pt,0);

\draw (P) circle (10pt);
\draw (Q) circle (10pt); 

\node[below] at (0,-10pt) {2n-1};
\node[below] at (40pt,-10pt) {2n};

\coordinate (P1) at (0,4pt);
\coordinate (P2) at (0,-4pt);
\coordinate (Q1) at (40pt,-4pt);
\coordinate (Q2) at (40pt,4pt);

\draw[black] (P1) circle (1.7pt);
\filldraw (P2) circle (1.7pt);
\draw[black] (Q1) circle (1.7pt);
\filldraw (Q2) circle (1.7pt);

\draw[-,out = 60,in = 120, looseness = 1.5] (P1) to (Q2);
\draw[-,out = -60,in = -120, looseness = 1.5] (P2) to (Q1);
\end{tikzpicture}
\end{matrix} = \frac{(2n|2)}{D^{2n}} \times \frac{1}{D^{2n-1}} \times D^2 \times D^{n-1} \times (1-S(t))^{2n-1}S(t).$$
This diagram has the same overall power of $D$, i.e., $D^{3n-2}$. Of course, we could have erased any of the legs from the $2n$ different legs in the fully contracted diagram, so the sum over all such diagrams gives
$$2\left(\begin{matrix}n \\ 1\end{matrix}\right)\frac{(2n|2)}{D^{3n-2}}  (1-S(t))^{2n-1}S(t). $$
Next, we can consider diagrams with two lines erased. For now, we can ignore the case where we erase both lines from the same pair of contracted blobs; this is because in this case we would be left with two blobs with no legs attached, and such diagrams only contribute to the Wigner function at $q=0$, but recall that we are currently looking at the Wigner function at $q\neq 0$. Thus, we need only consider the case where two legs are erased from two different pairs of contracted blobs. The sum over all such diagrams gives
$$4\left(\begin{matrix}n \\ 2\end{matrix}\right)\frac{(2n|2)}{D^{3n-2}}  (1-S(t))^{2n-2}S^2(t). $$  
Note that this diagram also contributes at the same order. Carrying on this way, we can erase $m$ legs from $m$ different pairs of contracted blobs to obtain:
$$\begin{matrix}
\begin{tikzpicture}
\coordinate (P) at (0,0);
\coordinate (Q) at (40pt,0);

\draw (P) circle (10pt);
\draw (Q) circle (10pt); 

\node[below] at (0,-10pt) {1};
\node[below] at (40pt,-10pt) {2};

\coordinate (P1) at (0,4pt);
\coordinate (P2) at (0,-4pt);
\coordinate (Q1) at (40pt,-4pt);
\coordinate (Q2) at (40pt,4pt);

\draw[black] (P1) circle (1.7pt);
\filldraw (P2) circle (1.7pt);
\draw[black] (Q1) circle (1.7pt);
\filldraw (Q2) circle (1.7pt);

\draw[-,out = -60,in = -120, looseness = 1.5] (P2) to (Q1);
\end{tikzpicture}\\
\begin{tikzpicture}
\filldraw (0,0) circle (0.5pt);
\filldraw (0,4pt) circle (0.5pt);
\filldraw (0,8pt) circle (0.5pt);
\end{tikzpicture}\\
\begin{tikzpicture}
\coordinate (P) at (0,0);
\coordinate (Q) at (40pt,0);

\draw (P) circle (10pt);
\draw (Q) circle (10pt); 

\node[below] at (0,-10pt) {2m-1};
\node[below] at (40pt,-10pt) {2m};

\coordinate (P1) at (0,4pt);
\coordinate (P2) at (0,-4pt);
\coordinate (Q1) at (40pt,-4pt);
\coordinate (Q2) at (40pt,4pt);

\draw[black] (P1) circle (1.7pt);
\filldraw (P2) circle (1.7pt);
\draw[black] (Q1) circle (1.7pt);
\filldraw (Q2) circle (1.7pt);

\draw[-,out = -60,in = -120, looseness = 1.5] (P2) to (Q1);
\end{tikzpicture}\\
\begin{tikzpicture}
\coordinate (P) at (0,0);
\coordinate (Q) at (40pt,0);

\draw (P) circle (10pt);
\draw (Q) circle (10pt); 

\node[below] at (0,-10pt) {2m+1};
\node[below] at (40pt,-10pt) {2m+2};

\coordinate (P1) at (0,4pt);
\coordinate (P2) at (0,-4pt);
\coordinate (Q1) at (40pt,-4pt);
\coordinate (Q2) at (40pt,4pt);

\draw[black] (P1) circle (1.7pt);
\filldraw (P2) circle (1.7pt);
\draw[black] (Q1) circle (1.7pt);
\filldraw (Q2) circle (1.7pt);

\draw[-,out = 60,in = 120, looseness = 1.5] (P1) to (Q2);
\draw[-,out = -60,in = -120, looseness = 1.5] (P2) to (Q1);
\end{tikzpicture}\\
\begin{tikzpicture}
\filldraw (0,0) circle (0.5pt);
\filldraw (0,4pt) circle (0.5pt);
\filldraw (0,8pt) circle (0.5pt);
\end{tikzpicture}\\
\begin{tikzpicture}
\coordinate (P) at (0,0);
\coordinate (Q) at (40pt,0);

\draw (P) circle (10pt);
\draw (Q) circle (10pt); 

\node[below] at (0,-10pt) {2n-1};
\node[below] at (40pt,-10pt) {2n};

\coordinate (P1) at (0,4pt);
\coordinate (P2) at (0,-4pt);
\coordinate (Q1) at (40pt,-4pt);
\coordinate (Q2) at (40pt,4pt);

\draw[black] (P1) circle (1.7pt);
\filldraw (P2) circle (1.7pt);
\draw[black] (Q1) circle (1.7pt);
\filldraw (Q2) circle (1.7pt);

\draw[-,out = 60,in = 120, looseness = 1.5] (P1) to (Q2);
\draw[-,out = -60,in = -120, looseness = 1.5] (P2) to (Q1);
\end{tikzpicture}
\end{matrix}
\quad =\quad \dfrac{(2n|2)}{D^{3n-2}}\begin{pmatrix}n\\m\end{pmatrix}(2S)^m(1-S)^{2n-m}.$$\\
Summing over all such diagrams for $m$ going from $0$ to $n$ (where the $m=0$ case corresponds to the fully contracted diagram), we get
\begin{eqnarray}
    I_n\Big|_{q\neq 0} &=& \sum_{m=0}^{n} \frac{(2n|2)}{D^{3n-2}}\begin{pmatrix}n\\m\end{pmatrix}(2S)^m(1-S)^{2n-m}\nonumber\\
    &=& \frac{(2n|2)}{D^{3n-2}}(1-S)^n\sum_{m=0}^{n} \begin{pmatrix}n\\m\end{pmatrix}(2S)^m(1-S)^{n-m}\nonumber\\
    &=& \frac{(2n|2)}{D^{3n-2}}(1-S)^n(1+S)^n\nonumber\\
     &=& \frac{(2n|2)}{D^{3n-2}}(1-S^2)^n.
\end{eqnarray}

\subsubsection*{Contribution from $q=0$}
For $q=0$, the leading contribution comes from the fully disconnected diagram (with no legs):
$$\begin{tikzpicture}
\coordinate (P) at (0,0);
\coordinate (Q) at (40pt,0);
\coordinate (R) at (80pt,0);
\coordinate (S) at (120pt,0);
\coordinate (T) at (200pt,0);

\draw (P) circle (10pt);
\draw (Q) circle (10pt); 
\draw (R) circle (10pt);
\draw (S) circle (10pt);
\draw (T) circle (10pt);

\node[below] at (0,-10pt) {1};
\node[below] at (40pt,-10pt) {2};
\node[below] at (80pt,-10pt) {3};
\node[below] at (120pt,-10pt) {4};
\node[below] at (200pt,-10pt) {2n};

\coordinate (P1) at (-4pt,0);
\coordinate (P2) at (4pt,0);
\coordinate (Q1) at (36pt,0);
\coordinate (Q2) at (44pt,0);
\coordinate (R1) at (76pt,0);
\coordinate (R2) at (84pt,0);
\coordinate (S1) at (116pt,0);
\coordinate (S2) at (124pt,0);
\coordinate (T1) at (196pt,0);
\coordinate (T2) at (204pt,0);

\draw[black] (P1) circle (1.7pt);
\filldraw (P2) circle (1.7pt);
\draw[black] (Q1) circle (1.7pt);
\filldraw (Q2) circle (1.7pt);
\draw[black] (R1) circle (1.7pt);
\filldraw (R2) circle (1.7pt);
\draw[black] (S1) circle (1.7pt);
\filldraw (S2) circle (1.7pt);

\filldraw (160pt,0) circle (1pt);
\filldraw (156pt,0) circle (1pt);
\filldraw (164pt,0) circle (1pt);

\draw[black] (T1) circle (1.7pt);
\filldraw (T2) circle (1.7pt);

\end{tikzpicture}$$\\
which gives 
\begin{align}
  I_n\biggr|_{q=0}= \frac{1}{D^{2n-1}}S^{2n}, 
\end{align}
where we get a $\frac{1}{D^{2n}}$ from the individual factors of $\frac{1}{D}$ in each copy of the Wigner function, and a factor of $D$ from the sum over $p$. Of course, there are more contributions at $q=0$; however, these contributions are sub-leading and turn out not to be important eventually. We will postpone a proper analysis of these other contributions to Appendix \ref{sec:irrelevant}. 

\begin{figure}[h!]
    \centering
\begin{tabular}{c c}   
    \includegraphics[width=0.5\linewidth]{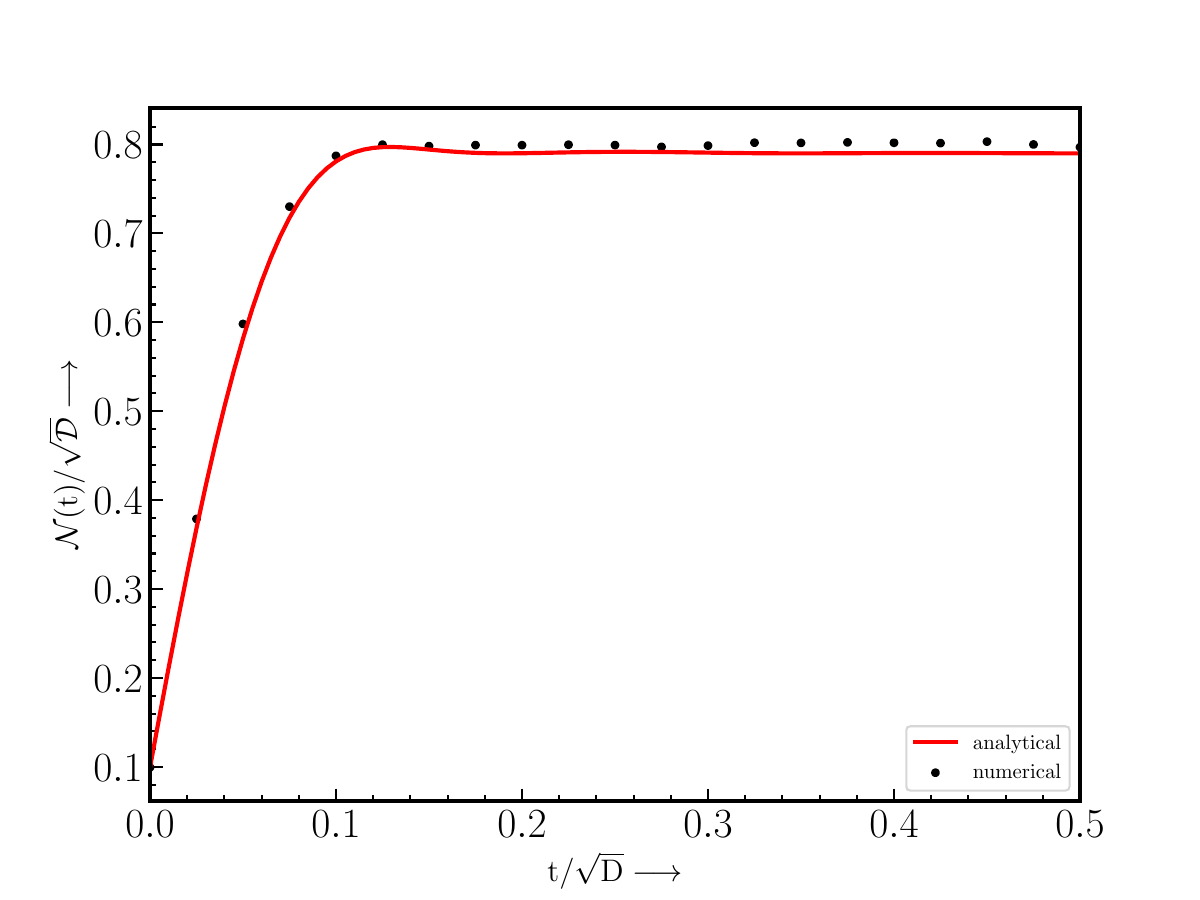} &
    \includegraphics[width=0.5\linewidth]{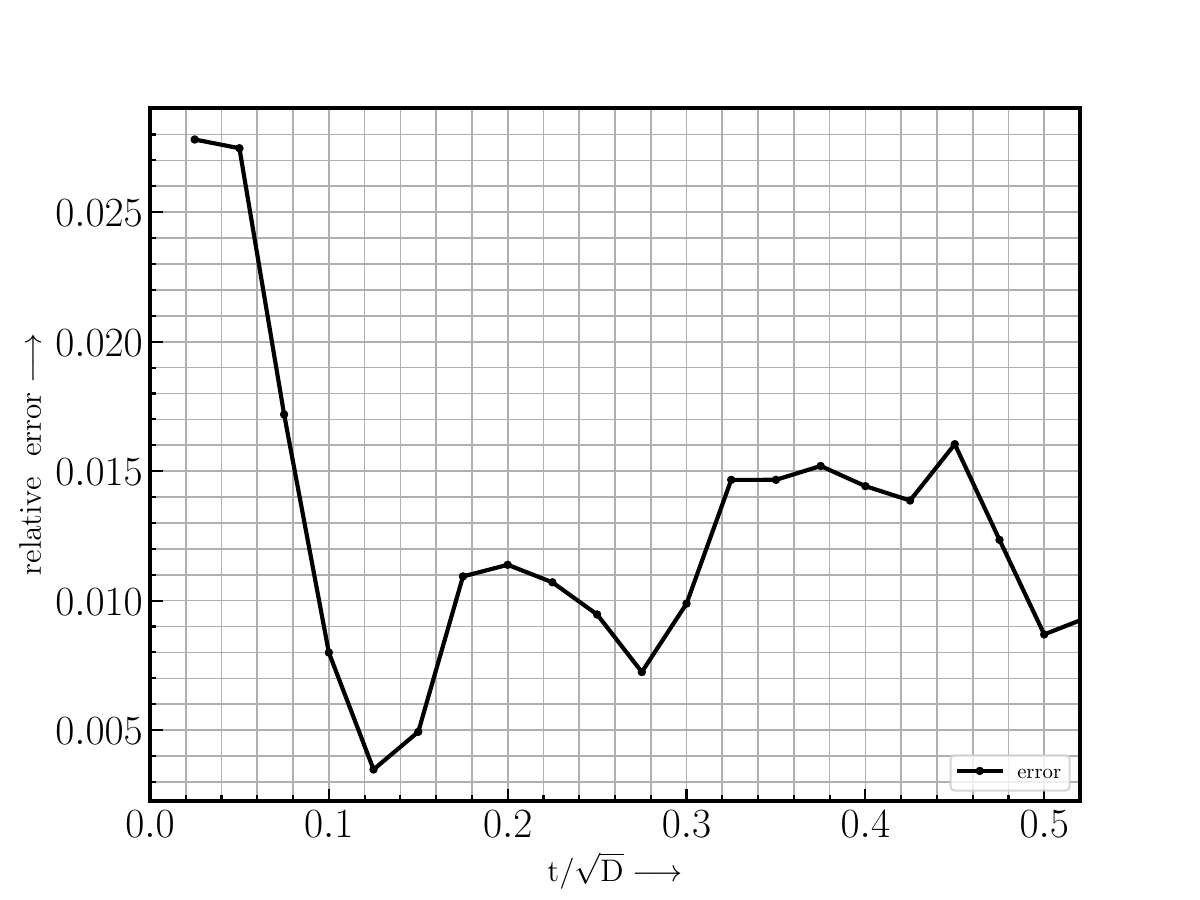}
    \end{tabular}
    \caption{The Wigner negativity in the computational basis for $D=101$: (left) the red curve shows the analytical formula in equation \eqref{avneg} for GUE with $\overline{S(t)}$ given by equation \eqref{SFFGUE}. The black points denote the numerically computed negativity for one specific choice of Hamiltonian drawn randomly from GUE. (right) The relative error $\frac{(\mathcal{N}_{\text{num.}}-\mathcal{N}_{\text{ana.}})}{\mathcal{N}_{\text{ana.}}}$ between the numerical and analytical formulas seems to be $O(1/D)$.}
  \label{fig:logloggeneric}
\end{figure}

Collecting together all the above contributions, we get
\begin{equation}
I_n = \frac{1}{D^{2n-1}}S^{2n} +  \frac{(2n|2)}{D^{3n-2}}(1-S^2)^n + \cdots,
\end{equation}
where $\cdots$ indicate lower order terms. In order to analytically continue this expression to $n=\frac{1}{2}$, we can write
$$ (2n|2) = \frac{(2n)!}{2^n n!} = \frac{\Gamma(2n+1)}{2^n \Gamma(n+1)}.$$
Upon analytic continuation, the combinatorial factor gives $\sqrt{\frac{2}{\pi}}$, and thus, we find that the Haar averaged Wigner negativity is given by
\beq 
I = S+ \sqrt{\frac{2D}{\pi}}(1-S^2)^{1/2}+\cdots.
\eeq 
The averaged Wigner negativity is now given by
\beq 
\overline{\mathcal{N}(\psi_t)}= \int dH\mu(H)\,I.
\eeq
Up to $O(1/\sqrt{D})$ corrections, we can thus write
\beq \label{avneg}
\overline{\mathcal{N}(\psi_t)} = \overline{S(t)}+ \sqrt{\frac{2D}{\pi}}(1-\overline{S(t)}^2)^{1/2}+\cdots.
\eeq
Equation \eqref{avneg} gives a universal formula for the averaged Wigner negativity for any unitarily invariant ensemble in terms of the averaged survival probability. From equation \eqref{SFF}, we see that $\overline{S(t)}$ starts off at one at $t=0$, but decays away from one in an $O(1)$ amount of time (see equation \eqref{SFFGUE} for the specific example of the GUE). Thus, we see from equation \eqref{avneg} that the negativity starts at one at $t=0$, but grows very rapidly and becomes $O(\sqrt{D})$ within an $O(1)$ amount of time evolution. In figure \ref{fig:logloggeneric}, we compare our analytical formula for the averaged negativity, equation \eqref{avneg}, with the numerically computed Wigner negativity for one specific Hamiltonian drawn from the GUE and find a match, up to $O(1/\sqrt{D})$ errors. So, to summarize, we have shown that the Wigner negativity in the computational basis -- or indeed any generic choice of basis with respect to which the Hamiltonian looks like a random matrix -- becomes exponentially large within an $O(1)$ amount of time evolution.

\section{Wigner negativity growth in the Krylov basis}
\label{sec:Krylov}

In this section, we will study the growth of Wigner negativity in the Krylov basis. Once again, in order to have some analytical handle on the problem, we will take the Hamiltonian to be randomly drawn from the GUE, and study the ensemble averaged negativity. In contrast to the previous section, we will see that the negativity in the Krylov basis grows much more gradually (as a power law) and becomes exponentially large only at exponential times. 

\subsection{Effective Hamiltonian and the large-$D$ limit}

Let us start with some initial state $|\psi_0\rangle$ and a Hamiltonian $H$ such that the initial state is generic with respect to the Hamiltonian (i.e., a random superposition of energy eigenstates). As mentioned above, we will take $H$ to be a random matrix drawn from GUE; then $\psi_0$ can be taken to be, say, the first computational basis state. Given the input $(\psi_0,H)$, we can construct a basis for the Hilbert space -- more precisely, for the subspace spanned by states of the form $e^{-itH}\psi_0$ -- called the Krylov basis. The construction goes as follows: take $|\psi_0\rangle$ to be the first basis vector. Then consider the state $H|\psi_0\rangle,$ and construct the second basis vector by orthogonalizing it with respect to $\psi_0$, and then properly normalizing the result. Then consider the state $H^2\psi_0$, and construct the third basis vector by orthogonalizing it with respect to the first two basis vectors and then normalizing the result, and so on. In other words, we consider the set of states $\left\{\psi_0, H\psi_0, H^2\psi_0\cdots\right\}$ and build an orthonormal basis from it using the Gram-Schmidt procedure. The resulting basis is called the Krylov basis. Note that the Krylov basis is an ordered basis; we will henceforth use the notation $|n\rangle$ to denote the $n$-th Krylov state. For the case of a random Hamiltonian acting on a generic initial state, we expect the time evolution to be ergodic, so that the Krylov basis spans the entire Hilbert space with $n$ ranging from $0$ to $D-1$.  

One of the key features of the Krylov basis is that the Hamiltonian is tri-diagonalized in this basis, i.e.,
\beq
H|n\rangle = a_n |n\rangle + b_{n+1}|n+1\rangle + b_n |n-1\rangle,
\eeq 
with the real coefficients $\{a_n\}_{n=0}^{D-1}$ and $\{b_n\}_{n=0}^{D-1}$ given by:
\beq 
a_n = \langle n| H|n\rangle, \;\;b_n = \langle n-1|H|n\rangle .
\eeq 
These numbers are called the Lanczos or Krylov coefficients. Remarkably, for Hamiltonians drawn randomly from the GUE, it was shown in \cite{10.1063/1.1507823, Balasubramanian:2022dnj} that the Krylov coefficients $\{a_n\}$ and $\{b_n\}$ behave as independent random variables with normal and chi distributions respectively. In the large-$D$ limit, their average values are given by
\beq \label{meanKCs}
\overline{a_n} = 0, \;\;\overline{b_n} = 1\;\;\; (n\;\text{fixed},D\to \infty),
\eeq 
with standard deviations of $O(1/\sqrt{D})$. Thus, in the large $D$ limit, we get an \emph{effective Hamiltonian} for dynamics in the fixed-$n$ regime:
\beq \label{Heff}
H_{\text{eff}}|n\rangle = |n+1\rangle + |n-1\rangle.
\eeq 
This effective Hamiltonian is sufficient to describe time evolution at sub-exponential times. A heuristic reason for this is as follows: consider the overlap $\langle n|e^{-itH}|0\rangle$. Series expanding the time evolution operator in $t$, we see that
\beq\label{TS}
\langle n|e^{-itH}|0\rangle =\sum_{M=0}^{\infty}\frac{t^M}{M!}\langle n|H^M|0\rangle = \sum_{M=0}^{\infty}\frac{t^M}{M!}\sum_{\gamma\in \text{paths}_M(0,n)}A(\gamma),
\eeq 
where $\text{paths}_M(0,n)$ is the set of all (discrete) paths from $0$ to $n$ on the lattice of Krylov indices consisting of a total number $M$ of unit time steps:
\beq 
\text{paths}_M(0,n) = \Big\{\gamma= \{n(s)\}\;\text{for}\;s\in (0,1,2,\cdots,M)\,|\;n(0)=0,\text{and}\;n(M)=n\Big\},
\eeq 
and $A(\gamma)$ is the amplitude associated to one such path $\gamma$:
\beq
A(\gamma) = \prod_{s=0}^M \langle n(s+1)|H|n(s)\rangle.
\eeq 
Since the Krylov basis tri-diagonalizes the Hamiltonian, we need only consider \emph{Motzkin paths} which comprise of either no jumps, or nearest-neighbor jumps -- the amplitude for such a path is simply the product of the corresponding Krylov coefficients $\{a_n\}$ and $\{b_n\}$ respectively. For any given $n$, the only terms which contribute in equation \eqref{TS} are ones where $M \geq n$; any term with $M < n$ will have a zero amplitude because the corresponding paths will not have sufficiently many time steps to reach the $n$th site. Now consider the case where $n$ and $t$ are fixed as $D\to \infty$. In this case, we expect that only paths which remain within the effective Hilbert space (i.e., the subspace with Krylov index fixed as $D\to \infty$) space dominate. The reason is that a path with $N$ time steps has a suppression factor of 
$$ \frac{t^N}{N!} \sim e^{-N( \log(N/t)-1)},$$ 
associated with it, where we have used the Stirling approximation for large $N$. If $N$ starts scaling with $D$, then at finite $t$ this leads to an exponential suppression in $D$ (or a doubly-exponential suppression in the number of degrees of freedom). This indicates that for sub-exponential times, the transition amplitude is dominated by paths which remain confined within the effective Hilbert space, where the effective Hamiltonian is given by equation \eqref{Heff}. Of course, the effective Hamiltonian ignores several properties of the Krylov coefficients: for instance, there are corrections to the average values of the Krylov coefficients away from equation \eqref{meanKCs} when $n$ becomes $O(D)$. There are also, of course, $O(\frac{1}{D})$ corrections coming from the statistical fluctuations of the Krylov coefficients. But since the dynamics is essentially confined to the effective Hilbert space at sub-exponential times, we do not expect either of these corrections to be important.    


The above heuristic analysis shows that for $O(1)$ times in the $D\to \infty$ limit, we can effectively treat the system as being infinite-dimensional, with the effective Hamiltonian given in equation \eqref{Heff}. This effective Hamiltonian can be interpreted as the quantization of a simple classical system -- the phase space is given by the semi-infinite cylinder $\mathcal{M}= \{(\ell,p) | \ell \geq 0, p \in S^1\} $. The classical Hamiltonian function is given by
\beq
H_{\text{cl}}(\ell, p) = 2\cos(p).
\eeq
In the quantum theory, the periodicity of $p$ forces quantization of $\ell$. It is an easy exercise to check that upon promoting the above function to an operator, the corresponding matrix elements in the position basis are given by equation \eqref{Heff}. The eigenstates of this effective Hamiltonian are well-known \cite{SusskindGlogower, phase} (see also Appendix E of \cite{Rabinovici:2023yex}):
\beq 
|\theta\rangle = \sqrt{\frac{2}{\pi}}\sum_{n=0}^{\infty}\sin[(n+1)\theta]|n\rangle,
\eeq
\beq 
H_{\text{eff}} |\theta\rangle =2\cos\,\theta |\theta\rangle.
\eeq 
Furthermore, these eigenstates satisfy the completeness relation:
\beq 
\int_0^{\pi}d\theta\,|\theta\rangle \langle \theta| = 1.
\eeq 
With this information at hand, we can compute the transition amplitude $\langle k | e^{-itH}|0\rangle$ in the large  $D$ limit:
\beqn\label{eq:aly}
\langle k|e^{-itH_{\text{eff}}}|0\rangle &=& \int_0^{\pi} d\theta\,\langle k|\theta\rangle\langle\theta |e^{-itH_{\text{eff}}}|0\rangle \nonumber\\
&=& \frac{2}{\pi}\int_0^{\pi} d\theta\,e^{-2it \cos\theta}\sin[(k+1)\theta]\sin(\theta)\nonumber\\
&=& \frac{1}{\pi}\int_0^{\pi} d\theta\,e^{-2it \cos\theta}\left(-\cos[(k+2)\theta]+\cos(k\theta)\right)\nonumber\\
&=&\frac{1}{i^k}\left(J_{k+2}(2t) + J_k(2t)\right)\nonumber\\
&=& \frac{1}{i^k}\frac{(k+1)}{t}J_{k+1}(2t),
\eeqn
where in going from the second to the third line, we have used the following integral representation for the Bessel function:
\beq \label{intrep}
J_{k}(z)= \frac{i^{-k}}{\pi}\int_0^zd\theta\,e^{iz\cos \theta}\cos(k\theta).
\eeq 
\begin{figure}[t]
    \centering
    \includegraphics[width=0.6\linewidth]{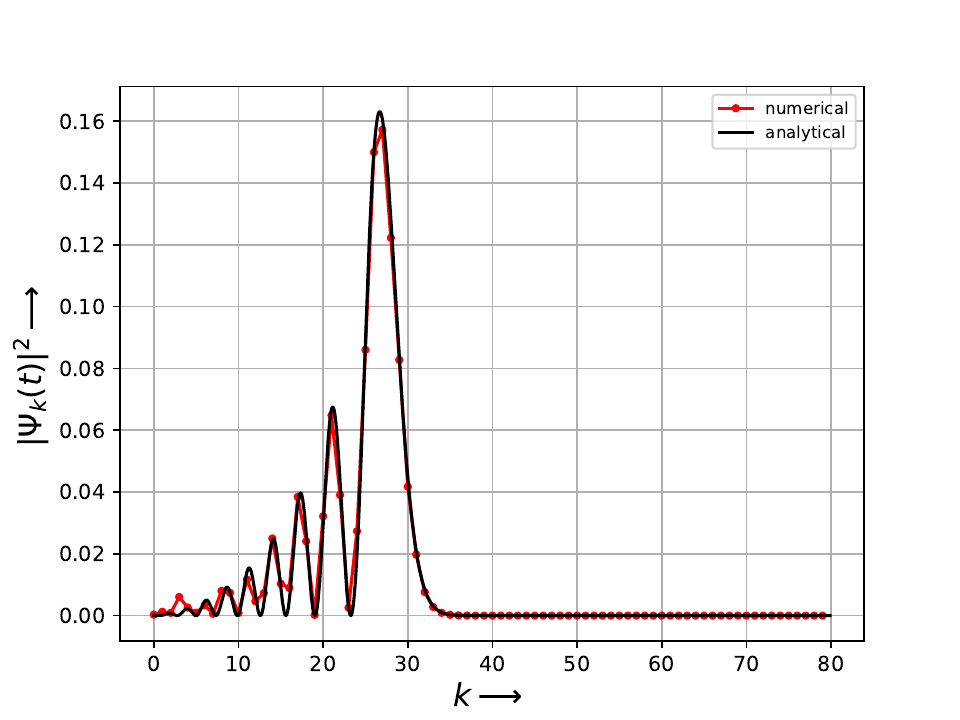}
    \caption{The probability as a function of $k$ for $t=15$ and $D=1499$. Numerically, the wave function ( $|\langle k | e^{-itH}|0\rangle|^2$ ) is calculated for a single Hamiltonian drawn from GUE. }
    \label{fig:WF}
\end{figure}

In figure \ref{fig:WF}, we compare the analytically computed transition probability (from equation \eqref{eq:aly}) obtained from the effective Hamiltonian at fixed $t$ and the numerically computed probability which uses the exact Hamiltonian, and find good agreement. This confirms our expectation that at finite $t$ the dynamics remains confined within the effective Hilbert space. Furthermore, we see that the wavefunction is oscillatory for $k\leq 2t$ and decays exponentially for $k>2t$. We can make this more precise in the large $t$-limit, $1 \ll t \ll D $. Said differently, we will be interested in the limits $D\to \infty$ followed by $t\to \infty,$ in that order. From the integral representation of the Bessel function \eqref{intrep}, we see that the $t \to \infty $ limit looks like a ``classical'' limit, where the exponential becomes very highly oscillatory. So, we can hope to do a stationary phase approximation. Of course, $k$ runs over non-negative integers, and so can itself scale with $t$. So, the natural limit to consider is $t\to \infty,\,k\to \infty$ with $x= \frac{k}{2t}$ fixed to be an arbitrary, non-negative real number. We can think of this as a ``continuum'' approximation of the wavefunction $\langle k|e^{-itH_{\text{eff}}}|0\rangle$, where instead of the discrete Krylov lattice, we now have a wavefunction of a continuous parameter $x$. In this limit, it is well-known that Bessel functions admit what is called the Debye expansion. In the range $0 < \frac{k}{z}<1$, we find the following oscillatory behavior:
\beq
J_{k}(z) \sim  \left(\frac{2}{\pi z \sin(\beta)}\right)^{1/2}\cos\left[z(\sin\,\beta - \beta \cos\,\beta)- \frac{\pi}{4}\right]\;\;\cdots \;\;(\cos\,\beta =\frac{k}{z}),
\eeq
while for the range $\frac{k}{z}>1$, the analogous formula shows that the Bessel function decays exponentially in $k$:

\beq
J_k(z) \sim \frac{1}{\sqrt{2 \pi z \sinh \alpha}} \exp\Big[-z(\alpha \cosh \alpha- \sinh \alpha )\Big]\;\; \cdots\;\; (\cosh\alpha= \frac{k}{z}).
\eeq
These formulas explain the qualitative features visible in figure \ref{fig:WF}, namely that the wavefunction is well-localized with the region $k \leq 2t$, and decays exponentially beyond this. This feature alone will turn out to be enough to show that the Wigner negativity cannot grow faster than $t$, as we will show next.



\subsection{Bound on negativity growth}

Our goal now is to use the above effective Hamiltonian description to bound the growth of Wigner negativity in the Krylov basis. The main point is that that the localized nature of the wavefunction in the Krylov basis (as discussed above) constrains the rate for negativity growth. An intuitive argument for this is as follows: we know that the negativity satisfies the upper bound
\beq \label{HU}
\log \mathcal{N}({\psi_t}) \leq H_{\frac{1}{2}}(Q),
\eeq 
where $H_{1/2}$ is the half-R\'enyi entropy of the probability distribution in the Krylov basis. As noted in the previous section, at a sufficiently large (but sub-exponential) time $t$, the wavefunction in the Krylov basis is essentially localized to the region $k \leq 2t,$ and decays exponentially beyond this. The half-R\'enyi entropy of any probability distribution localized within the region $k \leq 2t$ is bounded by
\beq \label{UD}
H_{\frac{1}{2}}(Q) \leq \log(2t),
\eeq 
where the inequality is saturated for the uniform distribution. From equations \eqref{HU} and \eqref{UD}, we see that we must have
\beq 
\mathcal{N}({\psi_t}) \leq 2t.
\eeq 
Thus, for large but sub-exponential times, we see that the negativity cannot grow faster than linear in time. In particular, this implies that the negativity remains sub-exponential for all sub-exponential times. 

The above argument was somewhat heuristic -- in particular, we did not account for the contribution to the R\'enyi entropy coming from the region $k > 2t$. Nevertheless, this argument captures the essence of why the negativity growth is polynomially bounded in the Krylov basis. We can give a slightly better bound (still polynomial) by using Jensen's inequality. Recall that the discrete Wigner function in the Krylov basis is given by:
\beq \label{KWan}
W(q,p) = \frac{1}{D}\sum_{k,\ell=0}^{D-1}\widehat{\delta}_{2q,k+\ell}e^{\frac{2\pi i}{D}(k-\ell)p} \langle k|e^{-itH}|0\rangle \langle 0| e^{itH}|\ell\rangle,
\eeq
where $|k\rangle$ and $|\ell\rangle$ now are Krylov basis vectors. In order to take $D\to \infty$, it is convenient to take $q$ to be a half integer ranging over $q=0,\frac{1}{2},1,\cdots, \frac{D-1}{2}$; this is just a re-labeling of the entries of the Wigner function. Taking the $D \to \infty$ limit and using the effective Hamiltonian description valid at $O(1)$ times leads to some important simplifications: firstly, we can take $q$ fixed (or more precisely $q/D \to 0$) in the $D\to \infty$ limit. Secondly, we can replace the $(\text{mod}\;D)$ Kronecker delta in the Wigner function with the ordinary Kronecker delta, essentially for the same reason as above. Finally, if we define 
\beq 
\theta = \frac{2\pi p}{D},
\eeq 
then we can treat the momentum as a continuous periodic variable. Thus, the effective phase space has a discrete position variable ranging over half-integer values and a continuous, period momentum variable. Defining 
\beq 
W(q,p) = \frac{2\pi}{D} W_{\text{eff}}(q,\theta = \frac{2\pi p}{D}),
\eeq 
we get the new effective Wigner function:
\beq \label{WigKry}
W_{\text{eff}}(q,\theta) = \frac{1}{2\pi}\sum_{k,\ell=0}^{\infty}\delta_{2q,k+\ell}e^{i(k-\ell)\theta} \psi_k^*(t) \psi_{\ell}(t).
\eeq 
In the large $D$ limit, the negativity of the Wigner function can now be written in terms of the effective Wigner function as
\beq 
\mathcal{N}(\psi_t) = \int \frac{d\theta}{2\pi} \sum_{q} |W_{\text{eff}}(q,\theta)|.
\eeq


Our goal now is to give a more precise bound on the negativity growth. At any given $t$, let us assume that the wavefunction has most of its support over $k \leq \Lambda_t$; for instance, in the present case of the Wigner function in the Krylov basis, we can take $\Lambda_t = 2xt$, for some positive number $x \gg 1$; the reason for introducing $x$ will become clear below. With this in mind, we can write the negativity as a sum of two terms:
\beq \label{neg}
\mathcal{N}(\psi_t) 
= \int \frac{d\theta}{2\pi} \left(\sum_{q=0}^{\Lambda_t}|W_{\text{eff}}(q,\theta)|+\sum_{q=\Lambda_t+\frac{1}{2}}^{\infty}|W_{\text{eff}}(q,\theta)|\right)
\eeq
The first term in \eqref{neg} can be bounded by using Jensen's inequality:
\beqn \label{bound1}
\int \frac{d\theta}{2\pi}\sum_{q=0}^{\Lambda_t}|W_{\text{eff}}(q,\theta)| &\leq &  \sqrt{2\Lambda_t+1} \left(\int \frac{d\theta}{2\pi}\sum_{q=0}^{\Lambda_t}|W_{\text{eff}}(q,\theta)|^2 \right)^{1/2}\nonumber\\ 
&\leq &  \sqrt{2\Lambda_t+1} \left(\int \frac{d\theta}{2\pi}\sum_{q=0}^{\infty}|W_{\text{eff}}(q,\theta)|^2 \right)^{1/2} = \sqrt{2\Lambda_t+1} .
\eeqn 
Since $\Lambda_t = 2xt$, this term cannot grow faster than $t^{1/2}$. 

\begin{figure}[t]
\centering
\includegraphics[height=9cm]{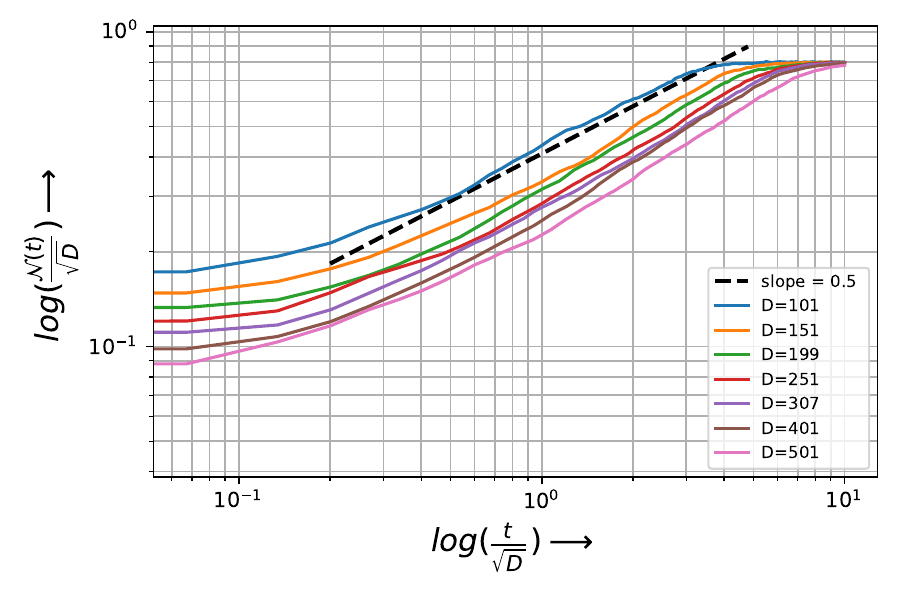}
\caption{$\log(\mathcal{N(\mathrm{t})})$ vs $\log(t)$ plot for different values of $D$. Here, the black dotted line represents the $\sqrt{t}$ curve.}
\label{fig:loglog sqrtt}
\end{figure}

The second term in equation \eqref{neg} involves the Wigner function in the region $q > \Lambda_t$. 
Here, we cannot use Jensen's inequality because the sum runs to infinity, but we can use the following argument: using the triangle inequality, we have
\beq
|W(q,p) | \leq \sum_{k,\ell=0}^{\infty}\delta_{2q,k+\ell}|\psi_k||\psi_{\ell}|.
\eeq 
Thus, the second term in \eqref{neg} obeys the following condition:
\beqn\label{WigBound}
\sum_{q > \Lambda_t} |W(q,p)|&\leq & \sum_{q > \Lambda_t}\sum_{k,\ell=0}^{\infty}\delta_{2q,k+\ell}|\psi_k||\psi_{\ell}|\nonumber\\
&=&\sum_{q> \Lambda_t}\left(\sum_{k,\ell>\Lambda_t}\delta_{2q,k+\ell}|\psi_k||\psi_{\ell}| + 2\sum_{k=0}^{\Lambda_t}\sum_{\ell > \Lambda_t} \delta_{2q,k+\ell}|\psi_k||\psi_{\ell}|\right)\nonumber\\
&=&\sum_{k,\ell >\Lambda_t}|\psi_k||\psi_{\ell}| + 2\sum_{q> \Lambda_t}\sum_{k=0}^{\Lambda_t}\sum_{\ell > \Lambda_t} \delta_{2q,k+\ell}|\psi_k||\psi_{\ell}|\nonumber\\
& \leq & \left(\sum_{k \geq \Lambda_t} |\psi_k|\right)^2 + 2\sum_{q >\Lambda_t}\sum_{k=0}^{\Lambda_t} |\psi_k||\psi_{2q-k}|.
\eeqn 
Let us define:
\beq 
C_t = \sum_{k > \Lambda_t} |\psi_k|,
\eeq 
which is essentially the contribution to the R\'enyi entropy coming from the region outside the cutoff $\Lambda_t$. The first term on the rigth hand side of \eqref{WigBound} is then simply $C_t^2$. Using $|\psi_k| \leq 1$, we can bound the second term as:

\beq 
 2\sum_{k=0}^{\Lambda_t}\sum_{q > \Lambda_t}|\psi_k| |\psi_{2Q-k}| \leq  2\sum_{k=0}^{\Lambda_t}\sum_{q > \Lambda_t} |\psi_{2Q-k}| \leq  2 \Lambda_t C_t.
\eeq
Thus, we get
\beq \label{bound2}
\sum_{q > \Lambda_t}|W(q,p)| \leq C_t^2 + 2\Lambda_t C_t.
\eeq 
Combining \eqref{bound1} with \eqref{bound2}, we obtain the following bound on the growth of negativity:
\beq \label{negbound}
\mathcal{N}(\psi_t) \leq \sqrt{2\Lambda_t + 1 } + C_t^2 + 2\Lambda_t C_t.
\eeq
Roughly speaking, the first term above comes from the fact that wavefunction is localized within the region $k \leq \Lambda_t$, while the second and third terms bound the contribution from the region $k > \Lambda_t$.  For the case at hand (i.e., Wigner negativity in the Krylov basis) with $\Lambda_t= 2 xt$, we can now estimate this latter contribution. In the region $k > \Lambda_t$, $\frac{k}{2t} > x \gg 1$. In the large $t$ limit, using the Debye approximation we get 
\beqn
|\psi_{k}(t) |
&\approx&  \left(\frac{\cosh\alpha_k}{\pi t \tanh\alpha_k}\right)^{1/2}e^{-2t\cosh \alpha_k (\alpha_k - \tanh\alpha_k)}\;\;\;\cdots\;\;\; (\cosh \alpha_k = k/2t)\nonumber\\
&=&  \left(\frac{k}{2\pi t^2 \tanh\alpha_k}\right)^{1/2}e^{-k (\alpha_k - \tanh\alpha_k)}.
\eeqn
This gives us an upper bound for $C_t$:
\beqn
C_t &=& \sum_{k> \Lambda_t} |\psi_k| \nonumber \\
&\approx & \sum_{k> \Lambda_t}\left(\frac{k}{2\pi t^2 \tanh\alpha_k}\right)^{1/2}e^{-k (\alpha_k - \tanh\alpha_k)}\nonumber\\
&\leq & \sum_{k> \Lambda_t}\left(\frac{k}{2\pi t^2 \tanh\alpha_*}\right)^{1/2}e^{-k (\alpha_* - 1)}\;\;\cdots \;\;(\cosh\alpha_* = x),\nonumber\\
&\leq & \left(\frac{1}{2\pi t^2 \tanh\alpha_*}\right)^{1/2}\sum_{k> \Lambda_t}k e^{-k (\alpha_* - 1)}\nonumber\\
&=& \left(\frac{1}{2\pi t^2 \tanh\alpha_*}\right)^{1/2}\frac{e^{-(\Lambda_t+1)(\alpha_*-1)}}{(1-e^{-(\alpha_*-1)})^2}\left(1+\Lambda_t(1 - e^{-(\alpha_*-1)})\right).
\label{sup}
\eeqn
Thus, $C_t$ is exponentially suppressed in the large-$t$ limit. From equation \eqref{negbound}, we then find the bound is mainly controlled by the first term, and is given by:
\beq 
\mathcal{N}(\psi_t) \leq \sqrt{4xt}. 
\eeq 
In figure \ref{fig:loglog sqrtt} we show the Wigner negativity computed numerically for Hamiltonians drawn randomly from the GUE for various values of $D$ on a log-log plot. Numerically, we see that the Wigner negativity growth follows a power law $t^{\gamma}$ at $O(1)$ times, with the coefficient $\gamma < \frac{1}{2}$. Beyond this, the numerical plot shows saturation to a value approximately equal to $\sqrt{\frac{2D}{\pi}}$, the same saturation value that we encountered in a generic basis. These features seemingly kick in at an exponential time scale, indicating breakdown of the semi-classical effective description. 

\section{Discussion}\label{sec:discussion}
The Wigner negativity gives a basis-dependent, operational measure of the complexity of simulating the time evolution of a quantum state on a classical computer. It is naturally related to other measures of complexity, such as the spread complexity, and provides a concrete notion of classicality for abstract quantum systems. In this work, we have studied the growth of Wigner negativity under time evolution with chaotic Hamiltonians for generic initial states. We showed that for a generic choice of basis, the negativity for an initially classical state grows rapidly and becomes exponentially large within an $O(1)$ amount of time evolution. On the other hand, in the Krylov basis, the Wigner negativity grows slowly, as a power law for a long time, and becomes exponentially large only at exponentially long times. This shows that the Krylov basis is well-suited for an effective, semi-classical description for chaotic quantum dynamics. Indeed, the effective Hamiltonian in the Krylov basis can be obtained from the quantization of a simple classical system, i.e., the semi-infinite cylinder with the Hamiltonian function $H_{\text{cl}}(q,p) = 2\cos(p)$, with $p$ being the periodic direction.\footnote{Similar Hamiltonians were encountered in the quantization of periodic-dilaton gravity theories in \cite{Blommaert:2024ydx, Blommaert:2025avl}.} This effective theory for the GUE agrees with $q\to 0$ limit of $q$-deformed JT gravity \cite{Berkooz:2018jqr, Berkooz:2022mfk, Lin:2022rbf, Jafferis:2022wez, Rabinovici:2023yex, Okuyama:2022szh, Okuyama:2023byh, Nandy:2024zcd,  Almheiri:2024xtw, Xu:2024gfm, Xu:2024hoc, Blommaert:2024ydx, Blommaert:2025avl, Miyaji:2025ucp}. In the AdS/CFT correspondence, the dual gravitational description gives a semi-classical handle on the strong coupling dynamics of the boundary theory. In some sense, gravity is an \emph{efficient, classical simulation} of boundary quantum dynamics. In the toy model setting of this paper, we see this idea playing out concretely.  

There are several future directions to explore: 
\begin{enumerate}

\item The choice of the initial state in our setup needs clarification. For the most part, we have taken the initial state to be random. The same results also apply had we taken the initial state to be the maximally entangled state on two copies of the system, with time evolution acting on one side. It would be interesting to explore other possibilities, and the role of the initial state in determining the emergent effective Hamiltonian. It would also be interesting to extend the results of this paper to more general ensembles of Hamiltonians; a potentially direct generalization could be in the case of the ensemble dual to $q$-deformed JT gravity \cite{Jafferis:2022wez}.\footnote{We thank Jiuci Xu for discussion on this point.} 

\item The Wigner negativity gives a lower bound on a certain entropic measure of spread complexity. Importantly, the Wigner negativity has a natural interpretation as a measure of the complexity of simulating the time evolution of a state on a classical computer; as we discussed, this can be made operationally precise in terms of the resource theory of stabilizer computation. In our opinion, this strengthens the interpretation of various notions of spread complexity as genuine complexity measures. It would be interesting to make this more concrete, and to study connections, if any, to circuit complexity \cite{susskind2014computational, Susskind:2018pmk, nielsen2005geometric, Nielsen_2006, Jefferson_2017, Chapman_2018, Balasubramanian:2019wgd, Balasubramanian:2021mxo, Craps:2023ivc}.

\item Finally, our results may have some bearing on the question of whether gravity should be thought of as being dual to one boundary theory or an ensemble average over boundary theories. From our point of view, it is natural to regard the Krylov basis\footnote{Say with respect to the maximally entangled initial state, but as we mentioned above, this point needs further clarification.} as a non-perturbative generalization of the geometric length basis of gravity \cite{Lin:2022rbf, Rabinovici:2023yex} (see \cite{2024JHEP...10..220I} for a different perspective). Every specific choice of Hamiltonian has a specific gravity dual, namely, the Hamiltonian expressed in the Krylov basis  -- the bulk-to-boundary map is merely the corresponding unitary change of basis.  Of course, for sub-exponential times, time evolution with all these Hamiltonians is well-described by the same universal, effective Hamiltonian, but microscopically, there is an ensemble of ``gravitational'' theories. For GUE, this is the ensemble of tri-diagonal matrices discovered by Dumitriu and Edelman \cite{10.1063/1.1507823, Balasubramanian:2022dnj}. It would be interesting to see if this bulk ensemble average can correctly capture the sum over topologies in gravity, or at least some discrete version of it, perhaps along the lines of \cite{Okuyama:2023kdo}. 

\end{enumerate}

\section*{Acknowledgments}
We would like to thank Vijay Balasubramanian, Pawel Caputa, Abhijit Gadde, Shiraz Minwalla, Harshit Rajgadia, Sandip Trivedi and Jiuci Xu for helpful discussions. This research was partly carried out at the International Center for Theoretical Sciences (ICTS) during the program ``Quantum Information, Quantum Field Theory and Gravity'' (code: ICTS/qftg2024/08). We acknowledge support from the Department of
Atomic Energy, Government of India, under project identification number RTI 4002, and from the Infosys Endowment for the study of the Quantum Structure of Spacetime.

\appendix

\section{Subleading corrections}\label{sec:irrelevant}
In the main text, we only considered the fully disconnected diagram for $q=0$. Here, we study all the other diagrams and show that their net contribution is sub-leading upon analytic continuation. Consider the fully pairwise contracted diagram, and now imagine removing lines from this diagram. When two lines are removed from the same blob we get
$$\begin{matrix}
\begin{tikzpicture}
\coordinate (P) at (0,0);
\coordinate (Q) at (40pt,0);

\draw (P) circle (10pt);
\draw (Q) circle (10pt); 

\node[below] at (0,-10pt) {1};
\node[below] at (40pt,-10pt) {2};

\coordinate (P1) at (0,4pt);
\coordinate (P2) at (0,-4pt);
\coordinate (Q1) at (40pt,-4pt);
\coordinate (Q2) at (40pt,4pt);

\draw[black] (P1) circle (1.7pt);
\filldraw (P2) circle (1.7pt);
\draw[black] (Q1) circle (1.7pt);
\filldraw (Q2) circle (1.7pt);

\end{tikzpicture}\\
\begin{tikzpicture}
\coordinate (P) at (0,0);
\coordinate (Q) at (40pt,0);

\draw (P) circle (10pt);
\draw (Q) circle (10pt); 

\node[below] at (0,-10pt) {3};
\node[below] at (40pt,-10pt) {4};

\coordinate (P1) at (0,4pt);
\coordinate (P2) at (0,-4pt);
\coordinate (Q1) at (40pt,-4pt);
\coordinate (Q2) at (40pt,4pt);

\draw[black] (P1) circle (1.7pt);
\filldraw (P2) circle (1.7pt);
\draw[black] (Q1) circle (1.7pt);
\filldraw (Q2) circle (1.7pt);

\draw[-,out = 60,in = 120, looseness = 1.5] (P1) to (Q2);
\draw[-,out = -60,in = -120, looseness = 1.5] (P2) to (Q1);
\end{tikzpicture}\\
\begin{tikzpicture}
\coordinate (P) at (0,0);
\coordinate (Q) at (40pt,0);

\draw (P) circle (10pt);
\draw (Q) circle (10pt); 

\node[below] at (0,-10pt) {5};
\node[below] at (40pt,-10pt) {6};

\coordinate (P1) at (0,4pt);
\coordinate (P2) at (0,-4pt);
\coordinate (Q1) at (40pt,-4pt);
\coordinate (Q2) at (40pt,4pt);

\draw[black] (P1) circle (1.7pt);
\filldraw (P2) circle (1.7pt);
\draw[black] (Q1) circle (1.7pt);
\filldraw (Q2) circle (1.7pt);

\draw[-,out = 60,in = 120, looseness = 1.5] (P1) to (Q2);
\draw[-,out = -60,in = -120, looseness = 1.5] (P2) to (Q1);
\end{tikzpicture}\\
\begin{tikzpicture}
\filldraw (0,0) circle (0.5pt);
\filldraw (0,4pt) circle (0.5pt);
\filldraw (0,8pt) circle (0.5pt);
\end{tikzpicture}\\
\begin{tikzpicture}
\coordinate (P) at (0,0);
\coordinate (Q) at (40pt,0);

\draw (P) circle (10pt);
\draw (Q) circle (10pt); 

\node[below] at (0,-10pt) {2n-1};
\node[below] at (40pt,-10pt) {2n};

\coordinate (P1) at (0,4pt);
\coordinate (P2) at (0,-4pt);
\coordinate (Q1) at (40pt,-4pt);
\coordinate (Q2) at (40pt,4pt);

\draw[black] (P1) circle (1.7pt);
\filldraw (P2) circle (1.7pt);
\draw[black] (Q1) circle (1.7pt);
\filldraw (Q2) circle (1.7pt);

\draw[-,out = 60,in = 120, looseness = 1.5] (P1) to (Q2);
\draw[-,out = -60,in = -120, looseness = 1.5] (P2) to (Q1);
\end{tikzpicture}
\end{matrix}$$\\
This gives us
\begin{align}
    &\underbrace{\frac{1}{D^{2n}}}_{\mathrm{prefactor}} \times \underbrace{\frac{1}{D^{2n-2}}}_{\mathrm{sum\ of\ Weingartens}} \times \underbrace{\sum_{q,p} D^{n-1}\delta_{q0}}_{\substack{\mathrm{contribution\ from\ the}\\ \mathrm{above\ diagram}}} \times \underbrace{(2n-2|2)\begin{pmatrix}2n\\2\end{pmatrix}}_{\substack{\mathrm{no\ of\ ways\ to\ remove}\\ \mathrm{2\ lines\ from\ the\ same\ pair}}} \nonumber \\
   &\hspace{6cm} \times \underbrace{(1-S)^{2n-2}}_{2n-2\ \mathrm{lines}} \times \underbrace{S^2}_{2\ \mathrm{remaining}}\nonumber \\
=& \frac{(2n-2|2)}{D^{3n-2}}\begin{pmatrix}2n\\2\end{pmatrix}(2S)^2(1-S)^{2n-2}
\end{align}
This particular diagram contributes at $\mathcal{O}(1/D^{3n-2})$. Similarly, when three lines are removed, such that 2 are removed from a pair and another from any other blob as\\
$$\begin{matrix}
\begin{tikzpicture}
\coordinate (P) at (0,0);
\coordinate (Q) at (40pt,0);

\draw (P) circle (10pt);
\draw (Q) circle (10pt); 

\node[below] at (0,-10pt) {1};
\node[below] at (40pt,-10pt) {2};

\coordinate (P1) at (0,4pt);
\coordinate (P2) at (0,-4pt);
\coordinate (Q1) at (40pt,-4pt);
\coordinate (Q2) at (40pt,4pt);

\draw[black] (P1) circle (1.7pt);
\filldraw (P2) circle (1.7pt);
\draw[black] (Q1) circle (1.7pt);
\filldraw (Q2) circle (1.7pt);

\end{tikzpicture}\\
\begin{tikzpicture}
\coordinate (P) at (0,0);
\coordinate (Q) at (40pt,0);

\draw (P) circle (10pt);
\draw (Q) circle (10pt); 

\node[below] at (0,-10pt) {3};
\node[below] at (40pt,-10pt) {4};

\coordinate (P1) at (0,4pt);
\coordinate (P2) at (0,-4pt);
\coordinate (Q1) at (40pt,-4pt);
\coordinate (Q2) at (40pt,4pt);

\draw[black] (P1) circle (1.7pt);
\filldraw (P2) circle (1.7pt);
\draw[black] (Q1) circle (1.7pt);
\filldraw (Q2) circle (1.7pt);

\draw[-,out = -60,in = -120, looseness = 1.5] (P2) to (Q1);
\end{tikzpicture}\\
\begin{tikzpicture}
\coordinate (P) at (0,0);
\coordinate (Q) at (40pt,0);

\draw (P) circle (10pt);
\draw (Q) circle (10pt); 

\node[below] at (0,-10pt) {5};
\node[below] at (40pt,-10pt) {6};

\coordinate (P1) at (0,4pt);
\coordinate (P2) at (0,-4pt);
\coordinate (Q1) at (40pt,-4pt);
\coordinate (Q2) at (40pt,4pt);

\draw[black] (P1) circle (1.7pt);
\filldraw (P2) circle (1.7pt);
\draw[black] (Q1) circle (1.7pt);
\filldraw (Q2) circle (1.7pt);

\draw[-,out = 60,in = 120, looseness = 1.5] (P1) to (Q2);
\draw[-,out = -60,in = -120, looseness = 1.5] (P2) to (Q1);
\end{tikzpicture}\\
\begin{tikzpicture}
\filldraw (0,0) circle (0.5pt);
\filldraw (0,4pt) circle (0.5pt);
\filldraw (0,8pt) circle (0.5pt);
\end{tikzpicture}\\
\begin{tikzpicture}
\coordinate (P) at (0,0);
\coordinate (Q) at (40pt,0);

\draw (P) circle (10pt);
\draw (Q) circle (10pt); 

\node[below] at (0,-10pt) {2n-1};
\node[below] at (40pt,-10pt) {2n};

\coordinate (P1) at (0,4pt);
\coordinate (P2) at (0,-4pt);
\coordinate (Q1) at (40pt,-4pt);
\coordinate (Q2) at (40pt,4pt);

\draw[black] (P1) circle (1.7pt);
\filldraw (P2) circle (1.7pt);
\draw[black] (Q1) circle (1.7pt);
\filldraw (Q2) circle (1.7pt);

\draw[-,out = 60,in = 120, looseness = 1.5] (P1) to (Q2);
\draw[-,out = -60,in = -120, looseness = 1.5] (P2) to (Q1);
\end{tikzpicture}
\end{matrix}$$\\
This diagram contributes to $ D^{n-2}\delta_{q0}\delta_{q\overline{0}}=0$.
Similary, whenever any odd number of lines are removed in this way, it is 0. Thus, a general term with $2m$ terms removed looks like
$$\begin{matrix}
\begin{tikzpicture}
\coordinate (P) at (0,0);
\coordinate (Q) at (40pt,0);
\coordinate (R) at (80pt,0);
\coordinate (S) at (120pt,0);
\coordinate (T) at (200pt,0);

\draw (P) circle (10pt);
\draw (Q) circle (10pt); 
\draw (R) circle (10pt);
\draw (S) circle (10pt);
\draw (T) circle (10pt);

\node[below] at (0,-10pt) {1};
\node[below] at (40pt,-10pt) {2};
\node[below] at (80pt,-10pt) {3};
\node[below] at (120pt,-10pt) {4};
\node[below] at (200pt,-10pt) {2m};

\coordinate (P1) at (-4pt,0);
\coordinate (P2) at (4pt,0);
\coordinate (Q1) at (36pt,0);
\coordinate (Q2) at (44pt,0);
\coordinate (R1) at (76pt,0);
\coordinate (R2) at (84pt,0);
\coordinate (S1) at (116pt,0);
\coordinate (S2) at (124pt,0);
\coordinate (T1) at (196pt,0);
\coordinate (T2) at (204pt,0);

\draw[black] (P1) circle (1.7pt);
\filldraw (P2) circle (1.7pt);
\draw[black] (Q1) circle (1.7pt);
\filldraw (Q2) circle (1.7pt);
\draw[black] (R1) circle (1.7pt);
\filldraw (R2) circle (1.7pt);
\draw[black] (S1) circle (1.7pt);
\filldraw (S2) circle (1.7pt);

\filldraw (160pt,0) circle (0.5pt);
\filldraw (156pt,0) circle (0.5pt);
\filldraw (164pt,0) circle (0.5pt);

\draw[black] (T1) circle (1.7pt);
\filldraw (T2) circle (1.7pt);

\end{tikzpicture}\\
\begin{tikzpicture}
\coordinate (P) at (0,0);
\coordinate (Q) at (40pt,0);

\draw (P) circle (10pt);
\draw (Q) circle (10pt); 

\node[below] at (0,-10pt) {2m+1};
\node[below] at (40pt,-10pt) {2m+2};

\coordinate (P1) at (0,4pt);
\coordinate (P2) at (0,-4pt);
\coordinate (Q1) at (40pt,-4pt);
\coordinate (Q2) at (40pt,4pt);

\draw[black] (P1) circle (1.7pt);
\filldraw (P2) circle (1.7pt);
\draw[black] (Q1) circle (1.7pt);
\filldraw (Q2) circle (1.7pt);

\draw[-,out = 60,in = 120, looseness = 1.5] (P1) to (Q2);
\draw[-,out = -60,in = -120, looseness = 1.5] (P2) to (Q1);
\end{tikzpicture}\\
\begin{tikzpicture}
\filldraw (0,0) circle (0.5pt);
\filldraw (0,4pt) circle (0.5pt);
\filldraw (0,8pt) circle (0.5pt);
\end{tikzpicture}\\
\begin{tikzpicture}
\coordinate (P) at (0,0);
\coordinate (Q) at (40pt,0);

\draw (P) circle (10pt);
\draw (Q) circle (10pt); 

\node[below] at (0,-10pt) {2n-1};
\node[below] at (40pt,-10pt) {2n};

\coordinate (P1) at (0,4pt);
\coordinate (P2) at (0,-4pt);
\coordinate (Q1) at (40pt,-4pt);
\coordinate (Q2) at (40pt,4pt);

\draw[black] (P1) circle (1.7pt);
\filldraw (P2) circle (1.7pt);
\draw[black] (Q1) circle (1.7pt);
\filldraw (Q2) circle (1.7pt);

\draw[-,out = 60,in = 120, looseness = 1.5] (P1) to (Q2);
\draw[-,out = -60,in = -120, looseness = 1.5] (P2) to (Q1);
\end{tikzpicture}
\end{matrix}$$
and gives $=\dfrac{(2n-2m|2)}{D^{3n-m-1}}\begin{pmatrix}2n\\2m\end{pmatrix}S^{2m}(1-S)^{2n-2m}$. The contribution from all such diagrams is given by:
\begin{align}
    \mathcal{N}(\psi)^{(2n)}\biggr|_{q=0}=&\sum_{m=1}^{n}\dfrac{(2n-2m|2)}{D^{3n-m-1}}\begin{pmatrix}2n\\2m\end{pmatrix}S^{2m}(1-S)^{2n-2m}.
\end{align}
 The case $m=n$ is same as the completely disconnected diagram which was dealt with in the main text, but we include it in the sum above for completeness.
In this form, the analytic continuation to $n\to \frac{1}{2}$ is difficult because $n$ appears in the upper limit of the summation. But we can process this sum further to make the analytic continuation more tractable. We begin by making the change of variable $m'=  n-m$:
\beq 
\mathcal{N}(\psi)^{(2n)}\biggr|_{q=0} = \frac{1}{D^{2n-1}}\sum_{m=0}^{n-1}\dfrac{(2m|2)}{D^{m}} \begin{pmatrix}2n\\2m\end{pmatrix}S^{2n-2m}(1-S)^{2m}.
\eeq 
This form of the series is better to work with because terms with larger $m$ are manifestly more suppressed at large $D$. Note that the combinatorial coefficient appearing above is given by
\beq 
(2m|2)\begin{pmatrix}2n\\2m\end{pmatrix} = \frac{\Gamma(2m+1)}{2^m \Gamma(m+1)\Gamma(2n-2m+1)}.
\eeq 
Since for all integer values of $m > n$, $\frac{1}{\Gamma(2n-2m+1)}$ vanishes, so we can extend the sum above to infinity, and subtract the $m=n$ term: 
\beq
\mathcal{N}(\psi)^{(2n)}\biggr|_{q=0} = \frac{1}{D^{2n-1}}\sum_{m=0}^{\infty}\dfrac{\Gamma(2n+1)}{\Gamma(m+1)\Gamma(2n-2m+1)} S^{2n-2m}\left(\frac{1-S}{\sqrt{2D}}\right)^{2m} - \frac{1}{D^{3n-1}} (2n|2)(1-S)^{2n}.
\eeq
If we now set $n\to \frac{1}{2}$ in the above expression, we get
\begin{align}
 \lim_{n\to\ \frac{1}{2}} \mathcal{N}(\psi)^{(2n)}\biggr|_{q=0}&=
\sum_{m=0}^{\infty}\dfrac{1}{\Gamma(m+1)\Gamma(2-2m)} S^{1-2m}\left(\frac{1-S}{\sqrt{2D}}\right)^{2m} - \frac{1}{\sqrt{D}} \sqrt{\frac{2}{\pi}}(1-S)\\
&=  S - \sqrt{\frac{2}{\pi D}}(1-S),
\end{align}
where only the $m=0$ term contributes because $\frac{1}{\Gamma(2-2m)}$ vanishes for $m \neq 0$. Thus, we see that the contribution from $q=0$ at large $D$ is dominated by the fully disconnected diagram.

\bibliographystyle{JHEP}
\bibliography{Reference_wigner.bib}

@article{Gopakumar:2003ns,
    author = "Gopakumar, Rajesh",
    title = "{From free fields to AdS}",
    eprint = "hep-th/0308184",
    archivePrefix = "arXiv",
    doi = "10.1103/PhysRevD.70.025009",
    journal = "Phys. Rev. D",
    volume = "70",
    pages = "025009",
    year = "2004"
}

@article{Balasubramanian:2024ghv,
    author = "Balasubramanian, Vijay and Das, Rathindra Nath and Erdmenger, Johanna and Xian, Zhuo-Yu",
    title = "{Chaos and integrability in triangular billiards}",
    eprint = "2407.11114",
    archivePrefix = "arXiv",
    primaryClass = "hep-th",
    doi = "10.1088/1742-5468/adba41",
    journal = "J. Stat. Mech.",
    volume = "2025",
    number = "3",
    pages = "033202",
    year = "2025"
}

@article{Nandy:2024htc,
    author = "Nandy, Pratik and Matsoukas-Roubeas, Apollonas S. and Mart\'\i{}nez-Azcona, Pablo and Dymarsky, Anatoly and del Campo, Adolfo",
    title = "{Quantum dynamics in Krylov space: Methods and applications}",
    eprint = "2405.09628",
    archivePrefix = "arXiv",
    primaryClass = "quant-ph",
    reportNumber = "RIKEN-iTHEMS-Report-24",
    doi = "10.1016/j.physrep.2025.05.001",
    journal = "Phys. Rept.",
    volume = "1125-1128",
    number = "June 18",
    pages = "1--82",
    year = "2025"
}

@article{Nandy:2024zcd,
    author = "Nandy, Pratik",
    title = "{Tridiagonal Hamiltonians modeling the density of states of the double-scaled SYK model}",
    eprint = "2410.07847",
    archivePrefix = "arXiv",
    primaryClass = "hep-th",
    reportNumber = "RIKEN-iTHEMS-Report-24",
    doi = "10.1007/JHEP01(2025)072",
    journal = "JHEP",
    volume = "01",
    pages = "072",
    year = "2025"
}

@article{Miyaji:2025ucp,
    author = "Miyaji, Masamichi and Mori, Soichiro and Okuyama, Kazumi",
    title = "{Finite $N$ Bulk Hilbert Space in ETH Matrix Model for double-scaled SYK}",
    eprint = "2505.13194",
    archivePrefix = "arXiv",
    primaryClass = "hep-th",
    reportNumber = "YITP 25-71",
    month = "5",
    year = "2025"
}

@article{Almheiri:2024xtw,
    author = "Almheiri, Ahmed and Goel, Akash and Hu, Xu-Yao",
    title = "{Quantum gravity of the Heisenberg algebra}",
    eprint = "2403.18333",
    archivePrefix = "arXiv",
    primaryClass = "hep-th",
    doi = "10.1007/JHEP08(2024)098",
    journal = "JHEP",
    volume = "08",
    pages = "098",
    year = "2024"
}

@article{Chandra:2022fwi,
    author = "Chandra, Jeevan and Hartman, Thomas",
    title = "{Coarse graining pure states in AdS/CFT}",
    eprint = "2206.03414",
    archivePrefix = "arXiv",
    primaryClass = "hep-th",
    doi = "10.1007/JHEP10(2023)030",
    journal = "JHEP",
    volume = "10",
    pages = "030",
    year = "2023"
}

@article{Streltsov_2017,
   title={Colloquium: Quantum coherence as a resource},
   volume={89},
   ISSN={1539-0756},
   url={http://dx.doi.org/10.1103/RevModPhys.89.041003},
   DOI={10.1103/revmodphys.89.041003},
   number={4},
   journal={Reviews of Modern Physics},
   publisher={American Physical Society (APS)},
   author={Streltsov, Alexander and Adesso, Gerardo and Plenio, Martin B.},
   year={2017},
   month=oct }

@article{upcoming,
    author = "Basu, Ritam and Parrikar, Onkar and Paul, Suprakash and Rajgadia, Harshit",
    title = "{Wigner negativity for evaporating black holes}",
    journal = "to appear"
}

@article{Howard:2014zwm,
    author = "Howard, Mark and Wallman, Joel and Veitch, Victor and Emerson, Joseph",
    title = "{Contextuality supplies the \textquoteleft{}magic\textquoteright{} for quantum computation}",
    doi = "10.1038/nature13460",
    journal = "Nature",
    volume = "510",
    number = "7505",
    pages = "351--355",
    year = "2014"
}

@article{Delfosse_2017,
   title={Equivalence between contextuality and negativity of the Wigner function for qudits},
   volume={19},
   ISSN={1367-2630},
   url={http://dx.doi.org/10.1088/1367-2630/aa8fe3},
   DOI={10.1088/1367-2630/aa8fe3},
   number={12},
   journal={New Journal of Physics},
   publisher={IOP Publishing},
   author={Delfosse, Nicolas and Okay, Cihan and Bermejo-Vega, Juan and Browne, Dan E and Raussendorf, Robert},
   year={2017},
   month=dec, pages={123024} }

@article{Leone2,
  title = {Stabilizer R\'enyi Entropy},
  author = {Leone, Lorenzo and Oliviero, Salvatore F. E. and Hamma, Alioscia},
  journal = {Phys. Rev. Lett.},
  volume = {128},
  issue = {5},
  pages = {050402},
  numpages = {5},
  year = {2022},
  month = {Feb},
  publisher = {American Physical Society},
  doi = {10.1103/PhysRevLett.128.050402},
  url = {https://link.aps.org/doi/10.1103/PhysRevLett.128.050402}
}

@article{Gopakumar:2004qb,
    author = "Gopakumar, Rajesh",
    title = "{From free fields to AdS. 2.}",
    eprint = "hep-th/0402063",
    archivePrefix = "arXiv",
    doi = "10.1103/PhysRevD.70.025010",
    journal = "Phys. Rev. D",
    volume = "70",
    pages = "025010",
    year = "2004"
}

@article{Gopakumar:2005fx,
    author = "Gopakumar, Rajesh",
    title = "{From free fields to AdS: III}",
    eprint = "hep-th/0504229",
    archivePrefix = "arXiv",
    doi = "10.1103/PhysRevD.72.066008",
    journal = "Phys. Rev. D",
    volume = "72",
    pages = "066008",
    year = "2005"
}

@article{Berkooz:2022mfk,
    author = "Berkooz, Micha and Isachenkov, Misha and Isachenkov, Mikhail and Narayan, Prithvi and Narovlansky, Vladimir",
    title = "{Quantum groups, non-commutative AdS$_{2}$, and chords in the double-scaled SYK model}",
    eprint = "2212.13668",
    archivePrefix = "arXiv",
    primaryClass = "hep-th",
    doi = "10.1007/JHEP08(2023)076",
    journal = "JHEP",
    volume = "08",
    pages = "076",
    year = "2023"
}

@article{Blommaert:2024ydx,
    author = "Blommaert, Andreas and Mertens, Thomas G. and Papalini, Jacopo",
    title = "{The dilaton gravity hologram of double-scaled SYK}",
    eprint = "2404.03535",
    archivePrefix = "arXiv",
    primaryClass = "hep-th",
    month = "4",
    year = "2024"
}

@article{Blommaert:2025avl,
    author = "Blommaert, Andreas and Levine, Adam and Mertens, Thomas G. and Papalini, Jacopo and Parmentier, Klaas",
    title = "{Wormholes, branes and finite matrices in sine dilaton gravity}",
    eprint = "2501.17091",
    archivePrefix = "arXiv",
    primaryClass = "hep-th",
    month = "1",
    year = "2025"
}

@article{Jafferis:2022wez,
    author = "Jafferis, Daniel Louis and Kolchmeyer, David K. and Mukhametzhanov, Baur and Sonner, Julian",
    title = "{Jackiw-Teitelboim gravity with matter, generalized eigenstate thermalization hypothesis, and random matrices}",
    eprint = "2209.02131",
    archivePrefix = "arXiv",
    primaryClass = "hep-th",
    doi = "10.1103/PhysRevD.108.066015",
    journal = "Phys. Rev. D",
    volume = "108",
    number = "6",
    pages = "066015",
    year = "2023"
}

@article{Xu:2024gfm,
    author = "Xu, Jiuci",
    title = "{On Chord Dynamics and Complexity Growth in Double-Scaled SYK}",
    eprint = "2411.04251",
    archivePrefix = "arXiv",
    primaryClass = "hep-th",
    month = "11",
    year = "2024"
}

@article{Xu:2024hoc,
    author = "Xu, Jiuci",
    title = "{Von Neumann Algebras in Double-Scaled SYK}",
    eprint = "2403.09021",
    archivePrefix = "arXiv",
    primaryClass = "hep-th",
    month = "3",
    year = "2024"
}

@article{Okuyama:2022szh,
    author = "Okuyama, Kazumi",
    title = "{Hartle-Hawking wavefunction in double scaled SYK}",
    eprint = "2212.09213",
    archivePrefix = "arXiv",
    primaryClass = "hep-th",
    doi = "10.1007/JHEP03(2023)152",
    journal = "JHEP",
    volume = "03",
    pages = "152",
    year = "2023"
}

@article{Baiguera:2025dkc,
    author = "Baiguera, Stefano and Balasubramanian, Vijay and Caputa, Pawel and Chapman, Shira and Haferkamp, Jonas and Heller, Michal P. and Halpern, Nicole Yunger",
    title = "{Quantum complexity in gravity, quantum field theory, and quantum information science}",
    eprint = "2503.10753",
    archivePrefix = "arXiv",
    primaryClass = "hep-th",
    month = "3",
    year = "2025"
}

@article{Okuyama:2023byh,
    author = "Okuyama, Kazumi",
    title = "{End of the world brane in double scaled SYK}",
    eprint = "2305.12674",
    archivePrefix = "arXiv",
    primaryClass = "hep-th",
    doi = "10.1007/JHEP08(2023)053",
    journal = "JHEP",
    volume = "08",
    pages = "053",
    year = "2023"
}

@article{Okuyama:2023kdo,
    author = "Okuyama, Kazumi",
    title = "{Discrete analogue of the Weil-Petersson volume in double scaled SYK}",
    eprint = "2306.15981",
    archivePrefix = "arXiv",
    primaryClass = "hep-th",
    doi = "10.1007/JHEP09(2023)133",
    journal = "JHEP",
    volume = "09",
    pages = "133",
    year = "2023"
}

@article{10.1063/1.1507823,
    author = {Dumitriu, Ioana and Edelman, Alan},
    title = {Matrix models for beta ensembles},
    journal = {Journal of Mathematical Physics},
    volume = {43},
    number = {11},
    pages = {5830-5847},
    year = {2002},
    month = {11},
    issn = {0022-2488},
    doi = {10.1063/1.1507823},
    url = {https://doi.org/10.1063/1.1507823},
    eprint = {https://pubs.aip.org/aip/jmp/article-pdf/43/11/5830/19137139/5830\_1\_online.pdf},
}

@article{Gopakumar:2011ev,
    author = "Gopakumar, Rajesh",
    title = "{What is the Simplest Gauge-String Duality?}",
    eprint = "1104.2386",
    archivePrefix = "arXiv",
    primaryClass = "hep-th",
    month = "4",
    year = "2011"
}

@article{Gopakumar:2022djw,
    author = "Gopakumar, Rajesh and Mazenc, Edward A.",
    title = "{Deriving the Simplest Gauge-String Duality -- I: Open-Closed-Open Triality}",
    eprint = "2212.05999",
    archivePrefix = "arXiv",
    primaryClass = "hep-th",
    month = "12",
    year = "2022"
}

@article{WOOTTERS19871,
title = {A Wigner-function formulation of finite-state quantum mechanics},
journal = {Annals of Physics},
volume = {176},
number = {1},
pages = {1-21},
year = {1987},
issn = {0003-4916},
doi = {https://doi.org/10.1016/0003-4916(87)90176-X},
url = {https://www.sciencedirect.com/science/article/pii/000349168790176X},
author = {William K Wootters}
}

@article{Maldacena:1997re,
    author = "Maldacena, Juan Martin",
    title = "{The Large $N$ limit of superconformal field theories and supergravity}",
    eprint = "hep-th/9711200",
    archivePrefix = "arXiv",
    reportNumber = "HUTP-97-A097, HUTP-98-A097",
    doi = "10.4310/ATMP.1998.v2.n2.a1",
    journal = "Adv. Theor. Math. Phys.",
    volume = "2",
    pages = "231--252",
    year = "1998"
}

@article{Witten:1998qj,
    author = "Witten, Edward",
    title = "{Anti de Sitter space and holography}",
    eprint = "hep-th/9802150",
    archivePrefix = "arXiv",
    reportNumber = "IASSNS-HEP-98-15",
    doi = "10.4310/ATMP.1998.v2.n2.a2",
    journal = "Adv. Theor. Math. Phys.",
    volume = "2",
    pages = "253--291",
    year = "1998"
}

@article{Veitch_2014,
doi = {10.1088/1367-2630/16/1/013009},
url = {https://dx.doi.org/10.1088/1367-2630/16/1/013009},
year = {2014},
month = {jan},
publisher = {IOP Publishing},
volume = {16},
number = {1},
pages = {013009},
author = {Victor Veitch and S A Hamed Mousavian and Daniel Gottesman and Joseph Emerson},
title = {The resource theory of stabilizer quantum computation},
journal = {New Journal of Physics}
}

@article{Mari_2012,
   title={Positive Wigner Functions Render Classical Simulation of Quantum Computation Efficient},
   volume={109},
   ISSN={1079-7114},
   url={http://dx.doi.org/10.1103/PhysRevLett.109.230503},
   DOI={10.1103/physrevlett.109.230503},
   number={23},
   journal={Physical Review Letters},
   publisher={American Physical Society (APS)},
   author={Mari, A. and Eisert, J.},
   year={2012},
   month=dec }

@misc{gottesman1998heisenberg,
      title={The Heisenberg Representation of Quantum Computers}, 
      author={Daniel Gottesman},
      year={1998},
      eprint={quant-ph/9807006},
      archivePrefix={arXiv},
      primaryClass={quant-ph}
}

@article{Aaronson:2004xuh,
    author = "Aaronson, Scott and Gottesman, Daniel",
    title = "{Improved simulation of stabilizer circuits}",
    eprint = "quant-ph/0406196",
    archivePrefix = "arXiv",
    doi = "10.1103/PhysRevA.70.052328",
    journal = "Phys. Rev. A",
    volume = "70",
    number = "5",
    pages = "052328",
    year = "2004"
}

@article{Mele2024introductiontohaar,
  doi = {10.22331/q-2024-05-08-1340},
  url = {https://doi.org/10.22331/q-2024-05-08-1340},
  title = {Introduction to {H}aar {M}easure {T}ools in {Q}uantum {I}nformation: {A} {B}eginner's {T}utorial},
  author = {Mele, Antonio Anna},
  journal = {{Quantum}},
  issn = {2521-327X},
  publisher = {{Verein zur F{\"{o}}rderung des Open Access Publizierens in den Quantenwissenschaften}},
  volume = {8},
  pages = {1340},
  month = may,
  year = {2024}
}

@article{Wigner:1932,
  title = {On the Quantum Correction For Thermodynamic Equilibrium},
  author = {Wigner, E.},
  journal = {Phys. Rev.},
  volume = {40},
  issue = {5},
  pages = {749--759},
  numpages = {0},
  year = {1932},
  month = {Jun},
  publisher = {American Physical Society},
  doi = {10.1103/PhysRev.40.749},
  url = {https://link.aps.org/doi/10.1103/PhysRev.40.749}
}

@article{Balasubramanian:2022tpr,
    author = "Balasubramanian, Vijay and Caputa, Pawel and Magan, Javier M. and Wu, Qingyue",
    title = "{Quantum chaos and the complexity of spread of states}",
    eprint = "2202.06957",
    archivePrefix = "arXiv",
    primaryClass = "hep-th",
    doi = "10.1103/PhysRevD.106.046007",
    journal = "Phys. Rev. D",
    volume = "106",
    number = "4",
    pages = "046007",
    year = "2022"
}

@misc{susskind2014computational,
      title={Computational Complexity and Black Hole Horizons}, 
      author={Leonard Susskind},
      year={2014},
      eprint={1402.5674},
      archivePrefix={arXiv},
      primaryClass={hep-th}
}

@inproceedings{Susskind:2018pmk,
    author = "Susskind, Leonard",
    title = "{Three Lectures on Complexity and Black Holes}",
    eprint = "1810.11563",
    archivePrefix = "arXiv",
    primaryClass = "hep-th",
    doi = "10.1007/978-3-030-45109-7",
    publisher = "Springer",
    series = "SpringerBriefs in Physics",
    month = "10",
    year = "2018"
}

@misc{nielsen2005geometric,
      title={A geometric approach to quantum circuit lower bounds}, 
      author={Michael A. Nielsen},
      year={2005},
      eprint={quant-ph/0502070},
      archivePrefix={arXiv},
      primaryClass={quant-ph}
}

@article{Nielsen_2006,
   title={Quantum Computation as Geometry},
   volume={311},
   ISSN={1095-9203},
   url={http://dx.doi.org/10.1126/science.1121541},
   DOI={10.1126/science.1121541},
   number={5764},
   journal={Science},
   publisher={American Association for the Advancement of Science (AAAS)},
   author={Nielsen, Michael A. and Dowling, Mark R. and Gu, Mile and Doherty, Andrew C.},
   year={2006},
   month=feb, pages={1133–1135} }

@article{Jefferson_2017,
   title={Circuit complexity in quantum field theory},
   volume={2017},
   ISSN={1029-8479},
   url={http://dx.doi.org/10.1007/JHEP10(2017)107},
   DOI={10.1007/jhep10(2017)107},
   number={10},
   journal={Journal of High Energy Physics},
   publisher={Springer Science and Business Media LLC},
   author={Jefferson, Robert A. and Myers, Robert C.},
   year={2017},
   month=oct }

@article{Chapman_2018,
   title={Toward a Definition of Complexity for Quantum Field Theory States},
   volume={120},
   ISSN={1079-7114},
   url={http://dx.doi.org/10.1103/PhysRevLett.120.121602},
   DOI={10.1103/physrevlett.120.121602},
   number={12},
   journal={Physical Review Letters},
   publisher={American Physical Society (APS)},
   author={Chapman, Shira and Heller, Michal P. and Marrochio, Hugo and Pastawski, Fernando},
   year={2018},
   month=mar }

@article{Balasubramanian:2019wgd,
    author = "Balasubramanian, Vijay and Decross, Matthew and Kar, Arjun and Parrikar, Onkar",
    title = "{Quantum Complexity of Time Evolution with Chaotic Hamiltonians}",
    eprint = "1905.05765",
    archivePrefix = "arXiv",
    primaryClass = "hep-th",
    doi = "10.1007/JHEP01(2020)134",
    journal = "JHEP",
    volume = "01",
    pages = "134",
    year = "2020"
}

@article{Balasubramanian:2021mxo,
    author = "Balasubramanian, Vijay and DeCross, Matthew and Kar, Arjun and Li, Yue (Cathy) and Parrikar, Onkar",
    title = "{Complexity growth in integrable and chaotic models}",
    eprint = "2101.02209",
    archivePrefix = "arXiv",
    primaryClass = "hep-th",
    doi = "10.1007/JHEP07(2021)011",
    journal = "JHEP",
    volume = "07",
    pages = "011",
    year = "2021"
}

@article{Berkooz:2018jqr,
    author = "Berkooz, Micha and Isachenkov, Mikhail and Narovlansky, Vladimir and Torrents, Genis",
    title = "{Towards a full solution of the large N double-scaled SYK model}",
    eprint = "1811.02584",
    archivePrefix = "arXiv",
    primaryClass = "hep-th",
    doi = "10.1007/JHEP03(2019)079",
    journal = "JHEP",
    volume = "03",
    pages = "079",
    year = "2019"
}

@article{Lin:2022rbf,
    author = "Lin, Henry W.",
    title = "{The bulk Hilbert space of double scaled SYK}",
    eprint = "2208.07032",
    archivePrefix = "arXiv",
    primaryClass = "hep-th",
    doi = "10.1007/JHEP11(2022)060",
    journal = "JHEP",
    volume = "11",
    pages = "060",
    year = "2022"
}

@article{Rabinovici:2023yex,
    author = "Rabinovici, E. and S\'anchez-Garrido, A. and Shir, R. and Sonner, J.",
    title = "{A bulk manifestation of Krylov complexity}",
    eprint = "2305.04355",
    archivePrefix = "arXiv",
    primaryClass = "hep-th",
    doi = "10.1007/JHEP08(2023)213",
    journal = "JHEP",
    volume = "08",
    pages = "213",
    year = "2023"
}

@article{Balasubramanian:2022dnj,
    author = "Balasubramanian, Vijay and Magan, Javier M. and Wu, Qingyue",
    title = "{Tridiagonalizing random matrices}",
    eprint = "2208.08452",
    archivePrefix = "arXiv",
    primaryClass = "hep-th",
    doi = "10.1103/PhysRevD.107.126001",
    journal = "Phys. Rev. D",
    volume = "107",
    number = "12",
    pages = "126001",
    year = "2023"
}

@article{Gross_2006,
   title={Hudson’s theorem for finite-dimensional quantum systems},
   volume={47},
   ISSN={1089-7658},
   url={http://dx.doi.org/10.1063/1.2393152},
   DOI={10.1063/1.2393152},
   number={12},
   journal={Journal of Mathematical Physics},
   publisher={AIP Publishing},
   author={Gross, D.},
   year={2006},
   month=dec }

@article{Bravyi_2005,
   title={Universal quantum computation with ideal Clifford gates and noisy ancillas},
   volume={71},
   ISSN={1094-1622},
   url={http://dx.doi.org/10.1103/PhysRevA.71.022316},
   DOI={10.1103/physreva.71.022316},
   number={2},
   journal={Physical Review A},
   publisher={American Physical Society (APS)},
   author={Bravyi, Sergey and Kitaev, Alexei},
   year={2005},
   month=feb }

@article{Parker:2018yvk,
    author = "Parker, Daniel E. and Cao, Xiangyu and Avdoshkin, Alexander and Scaffidi, Thomas and Altman, Ehud",
    title = "{A Universal Operator Growth Hypothesis}",
    eprint = "1812.08657",
    archivePrefix = "arXiv",
    primaryClass = "cond-mat.stat-mech",
    doi = "10.1103/PhysRevX.9.041017",
    journal = "Phys. Rev. X",
    volume = "9",
    number = "4",
    pages = "041017",
    year = "2019"
}

@article{Rabinovici_2021,
   title={Operator complexity: a journey to the edge of Krylov space},
   volume={2021},
   ISSN={1029-8479},
   url={http://dx.doi.org/10.1007/JHEP06(2021)062},
   DOI={10.1007/jhep06(2021)062},
   number={6},
   journal={Journal of High Energy Physics},
   publisher={Springer Science and Business Media LLC},
   author={Rabinovici, E. and Sánchez-Garrido, A. and Shir, R. and Sonner, J.},
   year={2021},
   month=jun }

@misc{caputa2021geometry,
      title={Geometry of Krylov Complexity}, 
      author={Pawel Caputa and Javier M. Magan and Dimitrios Patramanis},
      year={2021},
      eprint={2109.03824},
      archivePrefix={arXiv},
      primaryClass={hep-th}
}

@article{Goto,
  title = {Probing chaos by magic monotones},
  author = {Goto, Kanato and Nosaka, Tomoki and Nozaki, Masahiro},
  journal = {Phys. Rev. D},
  volume = {106},
  issue = {12},
  pages = {126009},
  numpages = {26},
  year = {2022},
  month = {Dec},
  publisher = {American Physical Society},
  doi = {10.1103/PhysRevD.106.126009},
  url = {https://link.aps.org/doi/10.1103/PhysRevD.106.126009}
}

@article{Leonard,
  title = {Quantum-State Tomography and Discrete Wigner Function},
  author = {Leonhardt, Ulf},
  journal = {Phys. Rev. Lett.},
  volume = {74},
  issue = {21},
  pages = {4101--4105},
  numpages = {0},
  year = {1995},
  month = {May},
  publisher = {American Physical Society},
  doi = {10.1103/PhysRevLett.74.4101},
  url = {https://link.aps.org/doi/10.1103/PhysRevLett.74.4101}
}

@article{sphere,
  title = {Discrete Moyal-type representations for a spin},
  author = {Heiss, Stephan and Weigert, Stefan},
  journal = {Phys. Rev. A},
  volume = {63},
  issue = {1},
  pages = {012105},
  numpages = {11},
  year = {2000},
  month = {Dec},
  publisher = {American Physical Society},
  doi = {10.1103/PhysRevA.63.012105},
  url = {https://link.aps.org/doi/10.1103/PhysRevA.63.012105}
}

@article{Galois,
   title={Discrete phase space based on finite fields},
   volume={70},
   ISSN={1094-1622},
   url={http://dx.doi.org/10.1103/PhysRevA.70.062101},
   DOI={10.1103/physreva.70.062101},
   number={6},
   journal={Physical Review A},
   publisher={American Physical Society (APS)},
   author={Gibbons, Kathleen S. and Hoffman, Matthew J. and Wootters, William K.},
   year={2004},
   month=dec }

@article{quantumcomp,
  title = {Quantum computers in phase space},
  author = {Miquel, C\'esar and Paz, Juan Pablo and Saraceno, Marcos},
  journal = {Phys. Rev. A},
  volume = {65},
  issue = {6},
  pages = {062309},
  numpages = {14},
  year = {2002},
  month = {Jun},
  publisher = {American Physical Society},
  doi = {10.1103/PhysRevA.65.062309},
  url = {https://link.aps.org/doi/10.1103/PhysRevA.65.062309}
}

@article{classicality,
  title = {Classicality in discrete Wigner functions},
  author = {Cormick, Cecilia and Galv\~ao, Ernesto F. and Gottesman, Daniel and Paz, Juan Pablo and Pittenger, Arthur O.},
  journal = {Phys. Rev. A},
  volume = {73},
  issue = {1},
  pages = {012301},
  numpages = {9},
  year = {2006},
  month = {Jan},
  publisher = {American Physical Society},
  doi = {10.1103/PhysRevA.73.012301},
  url = {https://link.aps.org/doi/10.1103/PhysRevA.73.012301}
}

@article{SusskindGlogower,
  title = {Quantum mechanical phase and time operator},
  author = {Susskind, Leonard and Glogower, Jonathan},
  journal = {Physics Physique Fizika},
  volume = {1},
  issue = {1},
  pages = {49--61},
  numpages = {13},
  year = {1964},
  month = {Jul},
  publisher = {American Physical Society},
  doi = {10.1103/PhysicsPhysiqueFizika.1.49},
  url = {https://link.aps.org/doi/10.1103/PhysicsPhysiqueFizika.1.49}
}

@article{Wang_2019,
   title={Quantifying the magic of quantum channels},
   volume={21},
   ISSN={1367-2630},
   url={http://dx.doi.org/10.1088/1367-2630/ab451d},
   DOI={10.1088/1367-2630/ab451d},
   number={10},
   journal={New Journal of Physics},
   publisher={IOP Publishing},
   author={Wang, Xin and Wilde, Mark M and Su, Yuan},
   year={2019},
   month=oct, pages={103002} }

@article{Pashayan_2015,
   title={Estimating Outcome Probabilities of Quantum Circuits Using Quasiprobabilities},
   volume={115},
   ISSN={1079-7114},
   url={http://dx.doi.org/10.1103/PhysRevLett.115.070501},
   DOI={10.1103/physrevlett.115.070501},
   number={7},
   journal={Physical Review Letters},
   publisher={American Physical Society (APS)},
   author={Pashayan, Hakop and Wallman, Joel J. and Bartlett, Stephen D.},
   year={2015},
   month=aug }

@article{Craps:2023ivc,
    author = "Craps, Ben and Evnin, Oleg and Pascuzzi, Gabriele",
    title = "{A relation between Krylov and Nielsen complexity}",
    eprint = "2311.18401",
    archivePrefix = "arXiv",
    primaryClass = "quant-ph",
    month = "11",
    year = "2023"
}

@article{Veitch_2012,
doi = {10.1088/1367-2630/14/11/113011},
url = {https://dx.doi.org/10.1088/1367-2630/14/11/113011},
year = {2012},
month = {nov},
publisher = {IOP Publishing},
volume = {14},
number = {11},
pages = {113011},
author = {Victor Veitch and Christopher Ferrie and David Gross and Joseph Emerson},
title = {Negative quasi-probability as a resource for quantum computation},
journal = {New Journal of Physics}
}

@article{phase,
  title = {Phase and Angle Variables in Quantum Mechanics},
  author = {CARRUTHERS, P. and NIETO, MICHAEL MARTIN},
  journal = {Rev. Mod. Phys.},
  volume = {40},
  issue = {2},
  pages = {411--440},
  numpages = {0},
  year = {1968},
  month = {Apr},
  publisher = {American Physical Society},
  doi = {10.1103/RevModPhys.40.411},
  url = {https://link.aps.org/doi/10.1103/RevModPhys.40.411}
}

@article{Leone2021quantumchaosis,
  doi = {10.22331/q-2021-05-04-453},
  url = {https://doi.org/10.22331/q-2021-05-04-453},
  title = {Quantum {C}haos is {Q}uantum},
  author = {Leone, Lorenzo and Oliviero, Salvatore F. E. and Zhou, You and Hamma, Alioscia},
  journal = {{Quantum}},
  issn = {2521-327X},
  publisher = {{Verein zur F{\"{o}}rderung des Open Access Publizierens in den Quantenwissenschaften}},
  volume = {5},
  pages = {453},
  month = may,
  year = {2021}
}

@article{Cotler:2016fpe,
    author = "Cotler, Jordan S. and Gur-Ari, Guy and Hanada, Masanori and Polchinski, Joseph and Saad, Phil and Shenker, Stephen H. and Stanford, Douglas and Streicher, Alexandre and Tezuka, Masaki",
    title = "{Black Holes and Random Matrices}",
    eprint = "1611.04650",
    archivePrefix = "arXiv",
    primaryClass = "hep-th",
    reportNumber = "SU-ITP-16-19, SU-ITP-16/19, YITP-16-124",
    doi = "10.1007/JHEP05(2017)118",
    journal = "JHEP",
    volume = "05",
    pages = "118",
    year = "2017",
    note = "[Erratum: JHEP 09, 002 (2018)]"
}

@ARTICLE{2024JHEP...10..220I,
       author = {{Iliesiu}, Luca V. and {Levine}, Adam and {Lin}, Henry W. and {Maxfield}, Henry and {Mezei}, M{\'a}rk},
        title = "{On the non-perturbative bulk Hilbert space of JT gravity}",
      journal = {Journal of High Energy Physics},
         year = 2024,
        month = oct,
       volume = {2024},
       number = {10},
          eid = {220},
        pages = {220},
          doi = {10.1007/JHEP10(2024)220},
archivePrefix = {arXiv},
       eprint = {2403.08696},
 primaryClass = {hep-th},
       adsurl = {https://ui.adsabs.harvard.edu/abs/2024JHEP...10..220I}
}

@ARTICLE{2024JHEP...05..264B,
       author = {{Basu}, Ritam and {Ganguly}, Anirban and {Nath}, Souparna and {Parrikar}, Onkar},
        title = "{Complexity growth and the Krylov-Wigner function}",
      journal = {Journal of High Energy Physics},
         year = 2024,
        month = may,
       volume = {2024},
       number = {5},
          eid = {264},
        pages = {264},
          doi = {10.1007/JHEP05(2024)264},
archivePrefix = {arXiv},
       eprint = {2402.13694},
 primaryClass = {hep-th},
       adsurl = {https://ui.adsabs.harvard.edu/abs/2024JHEP...05..264B}
}
\end{document}